\documentclass[12pt, a4paper]{article}
\usepackage{graphicx}
\usepackage{amssymb}
\usepackage{amsmath}
\usepackage{bm}
\usepackage{color}
\usepackage{cite}
\usepackage{epstopdf}            
\usepackage{epsfig}
\usepackage{mathrsfs}  
\usepackage{slashed}
\usepackage{enumerate}
            
\newcommand{\beq}{\begin{equation}}
\newcommand{\eeq}{\end{equation}}

\usepackage[colorlinks=true, linkcolor=black, citecolor=black,
urlcolor=black]{hyperref}

\begin{document}

\begin{titlepage}


\vskip 2cm
\begin{center}

{\Large
{\bf 
Top FCNC induced by a $\bm{Z'}$ boson
}
}

\vskip 2cm

Sungwoong Cho$^{1}$,
P. Ko$^{2}$,
Jungil Lee$^{1}$,
Yuji Omura$^{3}$,
and
Chaehyun Yu$^{1}$

\vskip 0.5cm

{\it $^{1}$
	Department of Physics, Korea University, Seoul 02841, Korea}\\ [3pt]
{\it $^2$School of Physics, KIAS, Seoul 02455, Korea
}\\[3pt]
{\it $^3$
Department of Physics, Kindai University, Higashi-Osaka, Osaka 577-8502, Japan}\\[3pt]

\vskip 0.5cm

\begin{abstract}
We consider a $Z'$ model in which the $Z'$ boson couples only to up and top
quarks. For a simple setup, one can consider a family-dependent $U(1)'$ symmetry,
under which only right-handed up-type quarks are charged.
After symmetry breaking, the right-handed quarks mix with each other,
and top flavor-changing neutral currents (FCNCs) mediated by the $Z'$ boson can be generated at tree level.
We take into account several processes at the parton level to probe the top FCNCs, and
find that the same-sign top quark pair production or the triple top quark production is  
the most capable of testing the top FCNCs.
We also consider a few non-FCNC processes to probe the $Z'$ boson, for example,
dijet production at the LHC.
Finally, we perform numerical analysis at detector level and
find that the same-sign top quark pair production provides
a much more stringent bound on the FCNC coupling than the other processes
in most of the parameter space.
\end{abstract}

\end{center}
\end{titlepage}

\section{Introduction}
\label{sec;intro}

The Standard Model (SM) has passed through stringent tests in various experiments.
It is, however, widely believed that the SM is not the end of the story and new physics beyond the SM 
will be probed at experiments, e.g., the LHC. 
So far, a lot of models have been suggested for new physics.
One of the simple extensions of the SM is to consider a new $U(1)'$ gauge symmetry.
Such a  $U(1)'$ gauge symmetry could be effectively induced at low energy from high-scale 
models like $E_6$ GUT models  (see, for example, Ref.~\cite{u1,leptophobic}). 
One of the distinctive signatures of $U(1)'$ models would be the production of a $Z'$ boson, 
which is a massive gauge boson of $U(1)'$ symmetry.  

One of the typical searches for the $Z'$ boson proceeds through production of the dilepton or dijet at the 
$Z'$ resonance. That $Z'$ boson decays into a dilepton or dijet, and it usually has flavor-independent couplings to fermions up to the differences among $U(1)'$ charges.
It is apparent that any processes in which the $Z'$ boson is involved could be a probe of the 
$U(1)'$ model. 
In this work, we take into account a more general case in which couplings of the $Z'$ boson
to fermions are flavor-dependent, and flavor-changing neutral currents (FCNCs) are 
induced by the $Z'$ boson at tree level after electroweak and $U(1)'$ symmetry breaking.
Then, we note that several processes via FCNCs could provide opportunities
to probe the $U(1)'$ model. Depending on $U(1)'$ charge assignment and the strength of FCNCs, 
flavor-violating processes may be more efficient to probe the $Z'$ model in comparison 
with the conventional  dilepton or dijet production.

In the SM, FCNCs  exist at the loop level owing to  the Glashow-Iliopoulos-Maiani (GIM) mechanism~\cite{GIM}.
In addition to the loop suppression factor, most of the FCNC observables usually have another 
suppression  factor due to the GIM mechanism.
Therefore, FCNCs are very rare in the SM and have been paid great attention to in the search for 
new physics beyond the SM.
In particular, mixing between a meson and antimeson is very sensitive to  $Z'$ interaction
with FCNCs relevant to quarks. Such FCNCs can contribute to the mixing at tree level
or via box diagrams depending on underlying models describing interactions. 

In the case of top FCNCs, they can contribute to  $D^0$-$\overline{D^0}$ mixing through a loop-level diagram, but both $t$-$c$-$X$ and $t$-$u$-$X$ FCNCs, where $X$ is a scalar or vector boson,
are required for the mixing.
Hence, if only one of $t$-$u$-$X$ and $t$-$c$-$X$ FCNCs exists,
it must be searched for  through FCNC  top decays or direct production of top quarks.
So far several observables have been proposed to study top FCNCs at colliders.
The search for top FCNCs in which the top quark couples to SM particles would provide
the most stringent bound on FCNC couplings in the top quark decay because the top quark is the 
heaviest particle in the SM and a lot of top quarks are produced at the LHC.
For example, current bounds on the branching ratio of the top quark decay
into a $u$ quark and $Z$ boson (or Higgs boson, $H$) at 95\% confidence level (CL) are 
\[
\textrm{BR}(t\to Z u)<2.4~(1.7) \times 10^{-4}  \ \ \ {\rm  and} \ \ \ 
\textrm{BR}(t\to h u)<4.7~(2.4) \times 10^{-3}
\] 
as given by the CMS (ATLAS) Collaboration~\cite{CMS:2017twu,Aaboud:2018nyl,Sirunyan:2017uae,Aaboud:2017mfd}.
 
However, if top FCNCs are generated from a new particle  heavier than the top quark, 
then top quark decays through FCNCs would be highly suppressed. 
Thus, one has to investigate the FCNCs through the production of the top quark.
In the case of top-Higgs FCNCs via $t$-$q$-$H$ ($q=u$ or $c$) couplings, Atwood {\it et al.}~\cite{Atwood:2013ica} discussed three processes,
$pp\to tt (\bar{t}\bar{t})$, $pp\to tj (\bar{t}j)$, and $pp\to tjH(\bar{t}jH)$, where $j$ denotes a jet, and 
found that the last process is the most capable of yielding the best upper bound on the top FCNCs. 
 
In this work, we consider top FCNCs, which are generated by a new gauge
boson, $Z'$, heavier than the top quark. In the search for FCNCs, 
we take into account four FCNC processes: same-sign top quark pair production, $pp\to tt$ $(\bar{t}\bar{t})$; single top quark production, $pp\to tj (\bar{t}j)$; radiative $Z'$ production, $pp\to t Z' j(\bar{t}Z' j)$; and triple top quark production, $pp \to tt\bar{t}$ $(t\bar{t}\bar{t})$.
We calculate cross sections for the four FCNC processes and compare them
with each other. We search for the best FCNC process that yields the strongest upper bound on the top FCNC coupling.
Since the FCNC couplings are related to non-FCNC couplings, 
such as $t$-$t$-$Z'$ coupling, we also consider non-FCNC processes,
dijet production, $pp\to jj$, top quark pair production, $pp\to t\bar{t}$, and four top quark production, $pp\to t\bar{t}t\bar{t}$,
to compare them with FCNC processes.
Furthermore, since the triple top quark production involves
both FCNC and non-FCNC couplings, the constraints from the non-FCNC
processes must be considered~\cite{Hou:2019gpn}.
 
This paper is organized as follows.
In Sec.~\ref{sec;setup}, we introduce an effective model which provides
FCNCs through a $Z'$ exchange. For an example of the UV complete model,
we consider a $U(1)'$ model in which only right-handed up-type quarks are charged.
In Sec.~\ref{sec;analysis}, we perform numerical analyses
for the flavor-violating processes as well as the flavor-conserving processes at the parton level.
Section~\ref{sec;simulation} is devoted to the numerical analysis
at detector level for the same-sign top quark pair production
and triple top quark production.
Finally, we summarize our results in Sec.~\ref{sec;summary}.

\section{Simple model}
\label{sec;setup}

In this section, we introduce a model that predicts FCNCs through a $Z'$ exchange.
The $Z'$ boson would be a gauge boson of an additional $U(1)'$ gauge symmetry
or, in a more general case, any larger gauge symmetry.
We assume that only right-handed up and top quarks couple to the $Z'$ boson in the quark mass eigenstates after electroweak and $U(1)'$ symmetry breaking.  Otherwise the models would be 
strongly constrained by $K^0 -\overline{K^0}$ and $B_{d(s)} - \overline{B}_{d(s)}$ mixings.

Then the relevant Lagrangian for the $Z'$ interaction with right-handed up-type quarks is given by
\begin{equation}
-{\cal L} \supset g_{uu}  \bar{u}_R \gamma^\mu u_R Z^\prime_\mu
+ g_{tt}  \bar{t}_R \gamma^\mu t_R Z^\prime_\mu
+ g_{ut}  \bar{u}_R \gamma^\mu t_R Z^\prime_\mu
+ g_{tu}  \bar{t}_R \gamma^\mu u_R Z^\prime_\mu,
\label{efflag}%
\end{equation}
where
$g_{ij}$ ($i,j=u$ or $t$) are  $Z'$ couplings to right-handed up-type quarks and can be 
expressed in terms of the coupling constant $g'$ of $U(1)'$ symmetry and mixing angles
among right-handed up-type quarks \cite{Ko:2011vd,Ko:2011di,Ko:2012ud,Ko:2012sv}.
For a simple example of the UV complete model, one may consider the chiral $U(1)'$ model, 
which was originally proposed to resolve the anomaly in the top quark forward-backward 
asymmetry at the Tevatron~\cite{Ko:2011vd,Ko:2011di,Ko:2012ud,Ko:2012sv}.
In this class of  models, only right-handed up-type quarks are charged under $U(1)'$ symmetry 
in a flavor-dependent manner, while  other fermion fields are uncharged or charged in a 
flavor-universal manner.  

For one of possible scenarios, one can assume that only right-handed top quark is 
charged under $U(1)'$ symmetry, i.e., $(u_R^{'} ,c_R^{'} ,t_R^{'} )=(0,0,1)$, in the interaction basis.\footnote{The primed and the unprimed fields denote the fields in the interaction and the mass 
eigenstates, respectively.}
After symmetry breaking, three right-handed quarks mix with each other,
and in the mass eigenstates, FCNCs mediated by the $Z'$ boson may be generated.
If we consider mixing between only right-handed up and top quarks,  
the $Z'$ couplings can be obtained as
\begin{equation}
g_{uu}=g' \sin^2\alpha,\quad
g_{ut}=-g_{tu}= g' \sin\alpha \cos\alpha,\quad
g_{tt}=g' \cos^2\alpha,
\label{guu}%
\end{equation}
where $\alpha$ is the mixing between $u$ and $t$ quarks.
For a small mixing angle, $\sin \alpha \ll 1$, one obtains the following hierarchy among
couplings: $g_{tt}\gg g_{ut} \gg g_{uu}$.
In the case of $\sin\alpha \sim \cos\alpha$,  all three couplings
become comparable to one another.
For charge assignment on right-handed up-type quarks, $(u_R^{'} ,c_R^{'} ,t_R^{'} )=(1,0,0)$, 
the hierarchy among couplings could be $g_{uu} \gg g_{ut} \gg g_{tt}$
when the mixing angle is small. In general, one must consider mixing among three right-handed up-type quarks. 
If we assume Eq.~(\ref{guu}), we obtain the  relation among couplings as
\begin{equation}
g_{uu} g_{tt} = g_{ut}^2 = g_{tu}^2,
\end{equation}
where only two of the four couplings are independent. 
However, this relation is valid only to mixing between two quarks in Eq.~(\ref{guu}).
If we consider the mixing among three right-handed up-type quarks,
the relation is invalid, and one must extend Eq.~(\ref{guu}) to the 
more general case including $g_{uc}$, $g_{ct}$, and $g_{cc}$
as shown in Ref.~\cite{Ko:2011di}.
In general, the coupling $g_{ij}$ is expressed in terms of the rotation
matirx, $R_u$, which diagonalizes the mass matrix of
the right-handed up-type quarks:
$g_{ij}= (R_u)_{ik} u_k (R_u)_{kj}^\dagger$, where
$u_k$ is the $U(1)^\prime$ charge assignment on the right-handed 
up-type quarks.
We note that $g_{ij}$ could be complex in principle, but we shall assume all $g_{ij}$ are real
for simplicity in this paper.

For $g_{uu}\ll g_{ut}$, we can ignore $Z'u\bar{u}$ interaction so that
the $Z'$ boson might not be searched for from dijet signals at colliders. Then, the FCNC processes via $Z'u\bar{t}$ ($Z't\bar{u}$) interaction may be the most sensitive to probe this class of $Z'$ model,
whereas for $g_{ut}\ll g_{uu}$, dijet production through $u\bar{u}\to Z'\to u\bar{u}$ would become 
a more sensitive process.

Before the study of the detailed phenomenology, let us discuss how to
satisfy the anomaly-free conditions. 
In our setup, the $U(1)'$ gauge symmetry is chiral, so 
extra fermions charged under $U(1)'$ are required to achieve the anomaly-free conditions.
When the $U(1)'$ charge assignment is generic,
some simple matter contents are discussed 
in Refs.~\cite{Ko:2011vd,Ko:2011di}.
In the case with $(u_R^{'} ,c_R^{'} ,t_R^{'})=(0,0,1)$, 
the anomaly-free $U(1)'$ gauge symmetry
can be achieved, introducing a SM-vectorlike fermion, $T$, whose charge of the SM gauge symmetry is the same as that of the right-handed top quark.	Defining the $U(1)'$ charges of left-handed and right-handed $T$ as $1$ and $0$, respectively, we obtain an anomaly-free $U(1)'$ model.

In addition, the flavor-dependent $U(1)'$ gauge symmetry often requires extra scalars,
since the symmetry forbids some Yukawa couplings and disturbs realization of the realistic
mass matrices for quarks. As is discussed in Refs.~\cite{Ko:2011vd,Ko:2011di,Ko:2012ud,Ko:2012sv},
a simple way is to introduce extra Higgs doublets charged under $U(1)'$.
The extra Higgs fields deviate the $\rho$ parameter from one at  tree level, while
they could relax the bound on the same-sign top signals in the collider experiments~\cite{Ko:2011vd,Ko:2011di,Ko:2012ud} and could explain the deviation in the semileptonic $B$ decay \cite{Ko:2012sv}.

In this paper, we do not discuss the detail of the Higgs sector and the extra fermions,
simply assuming that $Z'$ dominates over the Higgs contribution to the flavor-violating processes as well as flavor-conserving processes.
We also assume that the extra fermions are heavy, and
$Z'$ does not decay to the extra fermions and the extra fermions are irrelevant to $Z'$ physics.

\section{Numerical analysis at parton level}
\label{sec;analysis}%

In this section, we study several processes at the parton level with which one can search for
couplings in the Lagrangian~(\ref{efflag}).
We fix the center-of-momentum energy as $\sqrt{s}=13$ TeV.
The integrated luminosity $\int {\mathscr{L}} dt=137$ fb$^{-1}$ is
taken into account in numerical analysis.
We perform the numerical analysis at leading order in $\alpha_s$
by making use of MADGRAPH~\cite{Alwall:2011uj,Alwall:2014hca}
after implementing the simple model given in Eq.~(\ref{efflag}) into the generator.

For a reference value, we take the $Z'$ mass, $m_{Z'}=2$ TeV,
but we scan $m_{Z'}$ from 1 to 5 TeV in numerical analysis.
In top quark pair production and four top quark production, we also consider
lower $m_{Z'}$ mass than $1$ TeV.
Since $m_{Z'}> 2 m_t$ in the overall region, the $Z'$ boson can decay into top 
quarks. In this model, the decay channels to quarks are
$Z'\to u\bar{u}$, $t\bar{t}$, $u\bar{t}$, and $t\bar{u}$.
In principle, there could exist small mixing between the $Z'$ boson
and SM neutral bosons through kinetic mixing or one-loop corrections,
but it is ignored by assuming that the mixing is very small.
Hence, the $Z'$ boson is leptophobic since it does not couple to the SM leptons.
In a UV complete model, the $Z'$ boson can also decay into two scalar bosons or
SM gauge bosons, and the decay pattern depends on a given scalar potential. Therefore, the decay width, $\Gamma_{Z'}$, of the $Z'$ boson depends on parameters in the scalar potential and corresponding kinematics among scalar and gauge bosons. Here, we take $\Gamma_{Z'}/m_{Z'}=0.1$ for simplicity.

Since the $Z'$ boson couples to up and top quarks, we consider four
processes which could play important roles to probe top FCNCs:
(a) the same-sign top quark pair production, $p p \to t t (\bar{t}\bar{t})$;
(b) the single top quark production, $p p \to t j_u (\bar{t}j_u)$, where 
$j_u=u$ or $\bar{u}$; 
(c) the radiative $Z'$ production,
$p p \to t  j_u Z' (\bar{t} j_u Z')$;
and
(d) the triple top quark production, $p p \to t t \bar{t} (t\bar{t}\bar{t})$.
We also consider  three flavor-conserving processes, which may play important roles
in searching for couplings in Eq.~(\ref{efflag}):
(a) the dijet production, $p p \to Z' \to u\bar{u}$;
(b) the top quark pair production, $p p \to t \bar{t}$;
and
(c) the four top quark production, $p p \to t \bar{t} t \bar{t}$.
All processes are mutually connected and must be considered simultaneously
to find more detailed properties of the $Z'$ boson.
In Ref.~\cite{Cao:2019qrb}, some FCNC processes were taken into account 
simultaneously, but there are two different points from our analysis.
The model in that paper is based on an effective theory
in which the scale of new interactions is beyond the reach of near-future colliders.
Therefore, new interactions at current colliders are expressed in terms of 
dimension-6 operators. Another point which one must keep in mind is
that $CP$-conserving interactions could be generated by renormalization group
evolution even though they do not exist at the new interaction scale.
Then, for a more complete discussion, $CP$-conserving processes have to be
considered.

\subsection{Same-sign top quark pair production}

In this subsection, we consider the same-sign top quark pair production
induced by the $Z'$ boson, which proceeds through $pp \to tt (\bar{t}\bar{t})$.
The same-sign top quark pair production has been paid attention to as
a sensitive process searching for new physics because the SM background is 
very small~\cite{Berger:2011ua,Cao:2011ew,Ebadi:2018ueq}. 
In particular, it could reject a lot of new models, which 
might be a resolution of the anomaly in the top forward-backward asymmetry
at the Tevatron~\cite{Chatrchyan:2011dk}. In our model, this process can occur in
the $t$ channel through a $Z'$ exchange with $uu$ or $\bar{u}\bar{u}$
initial partons. Thus, only the $g_{ut}(=-g_{tu})$ coupling is involved in production amplitudes.
For a reference value $m_{Z'}=2$ TeV, we obtain cross sections
for the same-sign top quark pair production at the leading order in $\alpha_s$:
\begin{equation}
\sigma(pp\to tt)=1.67\, g_{ut}^4~\textrm{pb}, ~~~~
\sigma(pp\to \bar{t}\bar{t})=0.028 \, g_{ut}^4~\textrm{pb}.
\end{equation}
Note that production cross sections are proportional to $g_{ut}^4$.
For $g_{ut}=0.5$, the sum of production cross sections is about $0.11$ pb.
Because the SM backgrounds are very small in these processes,
one can get a strong constraint on $g_{ut}$~\cite{Atwood:2013xg}.
The upper bound on the cross section for the same-sign top quark
pair production is obtained as $\sigma(pp\to tt(\bar{t}\bar{t}))< 1.2 $ pb 
by the CMS Collaboration at 95\% CL with the integrated luminosity 
$35.9$ fb$^{-1}$ at $\sqrt{s}=13$ TeV~\cite{Sirunyan:2017uyt}, 
while the upper bound at ATLAS is $89$ fb at 95\% CL with the integrated luminosity 
$36.1$ fb$^{-1}$ at $\sqrt{s}=13$ TeV~\cite{Aaboud:2018xpj}.
Then the FCNC coupling is constrained as
$g_{ut}<0.92$ ($0.48$) for $m_{Z'}=2$ TeV by the CMS (ATLAS) bound.

\begin{figure}[tb]
\begin{center}
	\includegraphics[width=75mm]{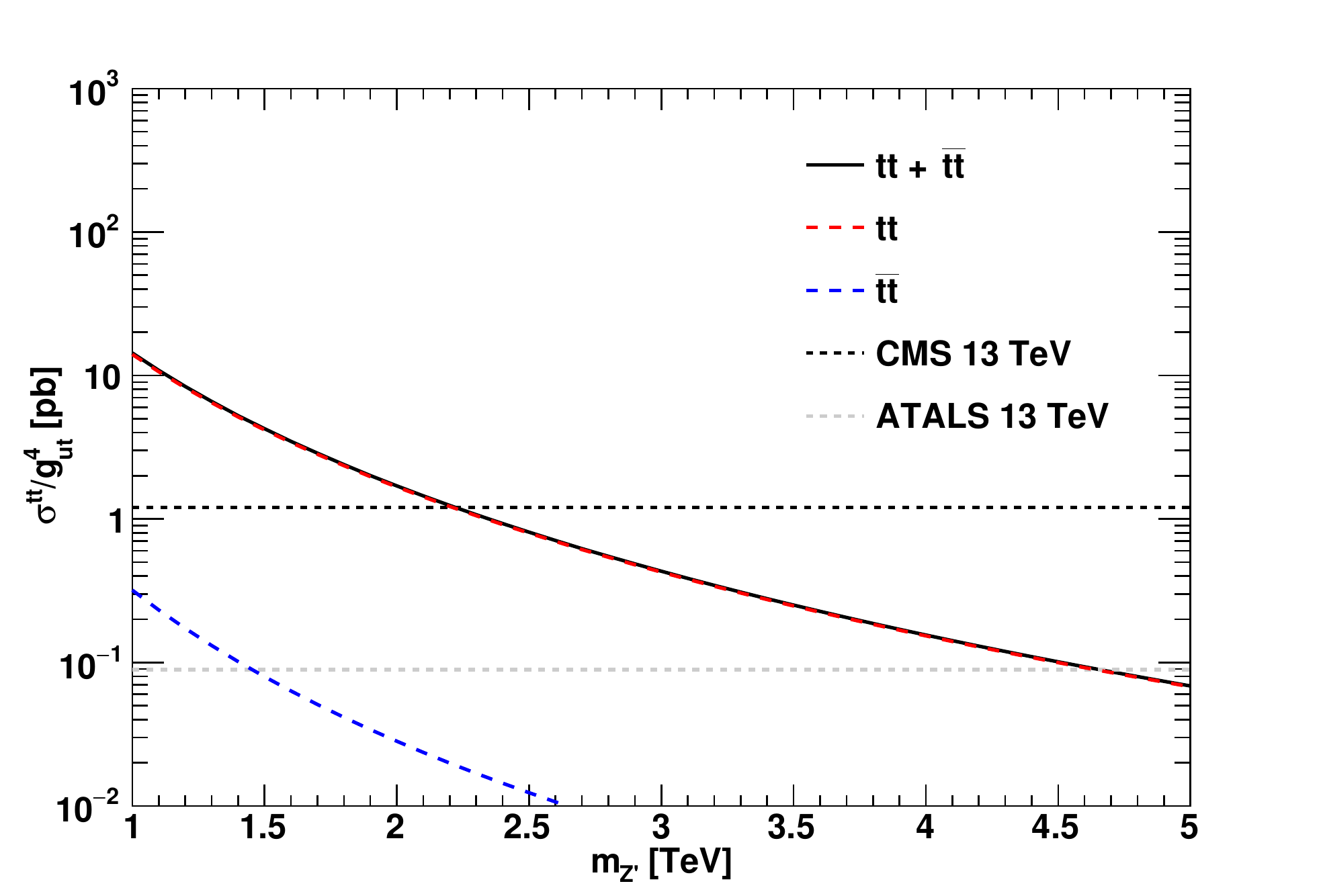} 
	\includegraphics[width=75mm]{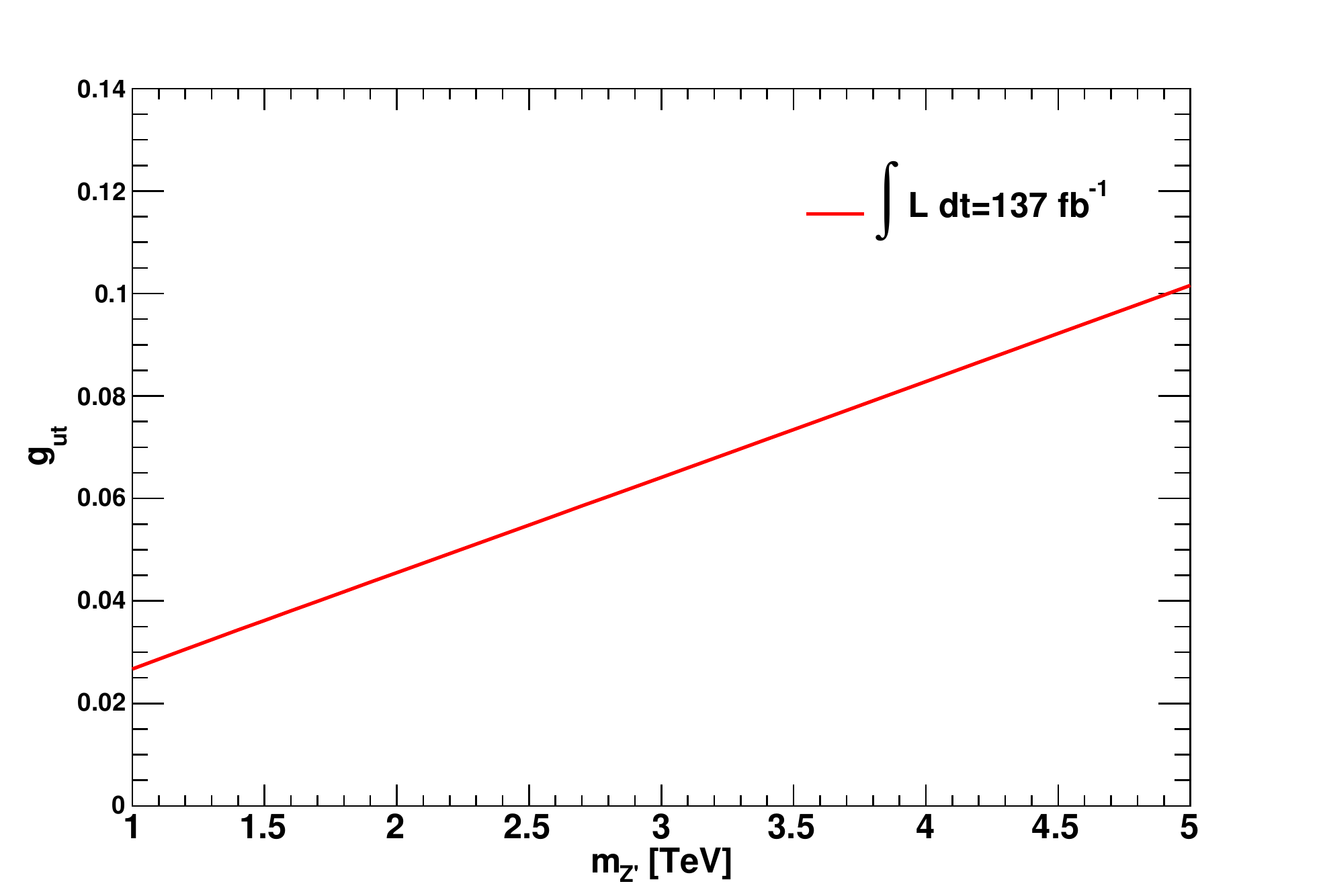} 
	\caption{ (Left) The cross section for the same-sign top quark pair production (in pb) divided by $g_{ut}^4$ as a function of $m_{Z'}$ (in TeV) at $\sqrt{s}=13$ TeV.
    (Right) The contour plot of $g_{ut}$ vs $m_{Z'}$ (in TeV) for 
    the number of signals, $S=1$, for  $\int{\mathscr L}dt=137$ fb$^{-1}$.}
	\label{fig:ss}
\end{center}
\end{figure}

In Fig.~\ref{fig:ss}, we show the results for the same-sign top quark pair
production. The left panel represents 
the total cross section for $pp\to tt$ (red line) and $pp\to \bar{t}\bar{t}$ (blue line) in units of pb 
divided by $g_{ut}^4$ as a function of the $Z'$ mass in units of TeV
at $\sqrt{s}=13$ TeV. The black line is the sum of them.
The horizontal lines indicate the upper bounds for
the same-sign top quark pair production at CMS (black dotted line)
and ATLAS (gray dashed line). 
For $g_{ut}=1$, the region of $m_{Z'} > 4.7$ TeV is allowed by the ATLAS bound and
for $m_{Z'}=1$ TeV, $g_{ut} > 0.28$ is excluded.
 Since the same-sign top quark pair production is forbidden at tree level in the SM, we 
find the region where the signal events do not exceed 1~\cite{Atwood:2013ica}.
The right panel of Fig.~\ref{fig:ss} represents
the contour plot of $g_{ut}$ vs $m_{Z'}$ (in TeV) for 
 the number of signals, $S=1$, for an integrated luminosity $\int{\mathscr L}dt=137$ fb$^{-1}$.
 Therefore, we expect that $g_{ut} \sim 0.046$ might be searched for
 through the same-sign top quark pair production 
 for an integrated luminosity of $137$ fb$^{-1}$ when $m_{Z'}=2$ TeV.
 Up to $m_{Z'}\sim 5$ TeV, the $g_{ut} \gtrsim 0.1$ region could be ruled out.
 However, this result is based on the analysis at the parton level.
 If one considers more realistic analyses or experimental uncertainties, 
 the bound could be approximately doubled~\cite{Atwood:2013ica} or greater~\cite{ATLASss,Aaboud:2018xpj}.

\begin{figure}[tb]
	\begin{center}
		\includegraphics[width=75mm]{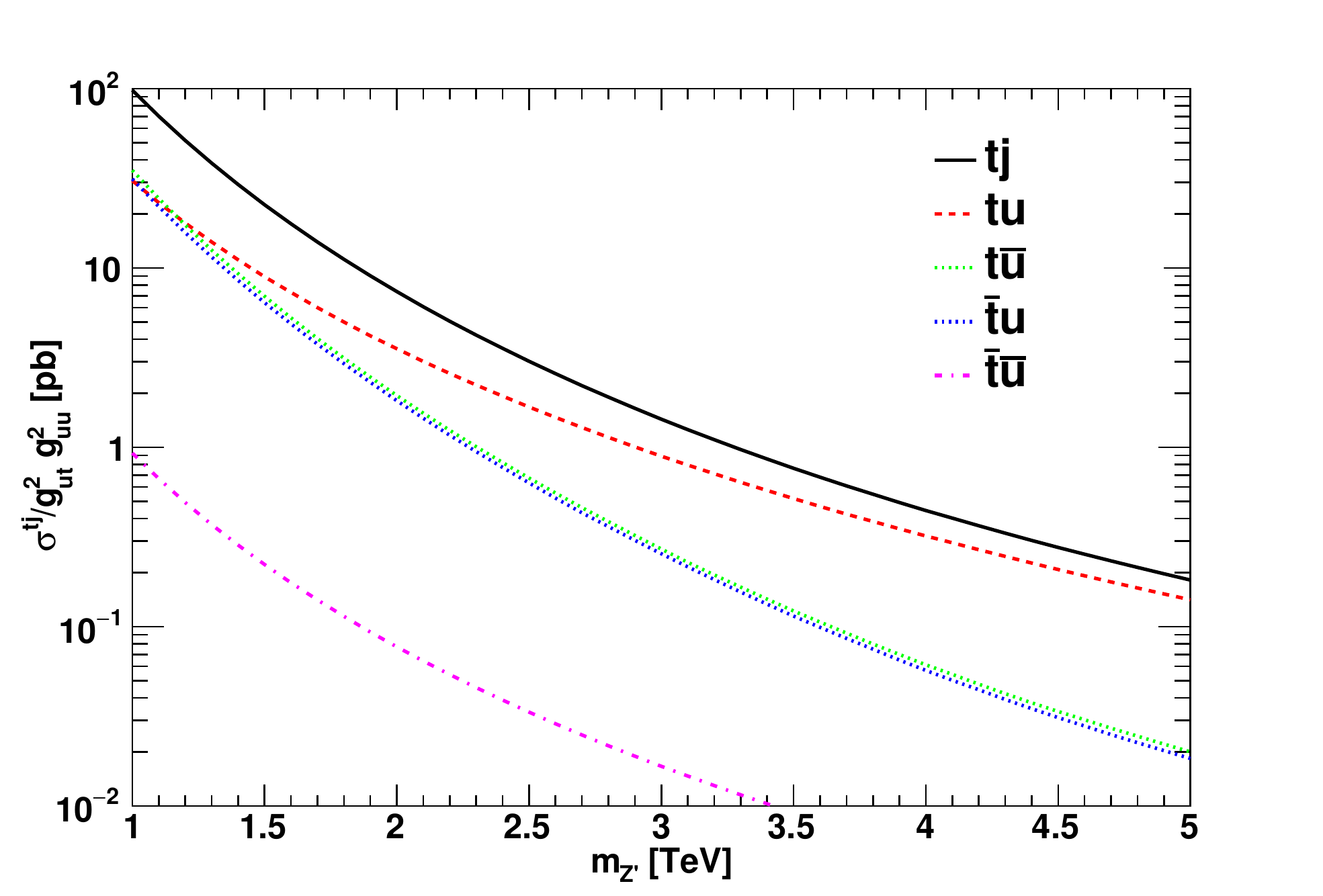} 
		\caption{ The cross sections for the single top quark production (in pb) divided by $g_{ut}^2 g_{uu}^2$ as a function of $m_{Z'}$ (in TeV) at $\sqrt{s}=13$ TeV.
			The red, blue, green, and purple lines correspond to $tu$, $t\bar{u}$, $\bar{t}u$, and $\bar{t}\bar{u}$ production, respectively, and the black
			line is the sum of them.	}
		\label{fig:tj}
	\end{center}
\end{figure}

\subsection{Single top quark production}

In this subsection, we consider the single top quark production in association 
with a $u$ or $\bar{u}$ quark. 
We note that four processes contribute to the single top quark production
mediated by the $Z'$ boson:
$pp\to tu(t\bar{u},\bar{t}u,\bar{t}\bar{u})$.

In the production of $tu$ or $\bar{t}\bar{u}$, there exists only one $t$-channel diagram mediated by the $Z'$ boson,
while both $s$- and $t$-channel diagrams via a $Z'$ exchange contribute to the
$t\bar{u}$ or $\bar{t}u$ production.
In each process, one of the vertices with which the $Z'$ boson is involved
contains $g_{ut}$ (or $g_{tu}$), while the other vertex contains $g_{uu}$.
Therefore, the cross section for the single top quark production is proportional to $g_{uu}^2 g_{ut}^2$.
For $m_{Z'}=2$ TeV, we calculate the cross sections for each process at $\sqrt{s}=13$ TeV,
\begin{subequations}
\begin{eqnarray}
\label{tta}
\sigma(pp\to t\bar{u}) &=& 1.94\, g_{ut}^2 g_{uu}^2~\textrm{pb}, \\
\sigma(pp\to tu) &=& 3.54 \,g_{ut}^2 g_{uu}^2~\textrm{pb}, \\
\sigma(pp\to \bar{t}u) &=& 1.82 \,g_{ut}^2 g_{uu}^2~\textrm{pb}, \\
\label{ttd}
\sigma(pp\to \bar{t}\bar{u}) &=& 0.077\, g_{ut}^2 g_{uu}^2~\textrm{pb},
\end{eqnarray}
\end{subequations}
where we implement the cuts on the associated $u$ (or $\bar{u}$) quark as
its transverse momentum $p_T \ge 30~\textrm{GeV}$ and 
rapidity $|\eta|<2.7$.
Because the parton density of the $u$ quark is greater than that of the $\bar{u}$ quark
inside the proton,
the cross section for the $tu$ production, whose parton process is $uu\to tu$, is the greatest among the four processes in Eqs.~(\ref{tta})--(\ref{ttd}).
The small difference between the production cross sections for $t\bar{u}$
and $\bar{t}u$  is due to the kinematic cuts on jets.
Since an incident $u$ quark would be more energetic than an incident $\bar{u}$ quark,
the $u$ quark in the final state would be faster than the $\bar{u}$ quark 
in the final state statistically. Then, more $u$ quarks would be removed
by the cuts on jets and the $t\bar{u}$ production would have a slightly larger cross section
than the $\bar{t}u$ production.
For $m_{Z'}=2$ TeV and $g_{ut}=g_{uu}=0.5$, the sum of the cross sections for the single top quark
production via a $Z'$ exchange is about $0.46$ pb. 

In Fig.~\ref{fig:tj}, we draw the cross sections for the single top quark production (in pb) divided by $g_{ut}^2 g_{uu}^2$ as a function of $m_{Z'}$ (in TeV) at $\sqrt{s}=13$ TeV.
The red, blue, green, and purple lines correspond to $tu$, $t\bar{u}$,
$\bar{t}u$, and $\bar{t}\bar{u}$ production, respectively.
Their sum is represented by the black line.
The $\bar{t}\bar{u}$ production is suppressed in the overall region
due to small incident parton density of the $\bar{u}$ quark in comparison with that
of the $u$ quark.
For $m_{Z'}\approx 1$ TeV, the production cross sections for $t\bar{u}$ or $\bar{t}u$ are comparable to that of $tu$.
However, for a larger value of $m_{Z'}$, the $tu$ production process dominates over
the other single top quark production processes.

\begin{figure}[tb]
	\begin{center}
		\includegraphics[width=75mm]{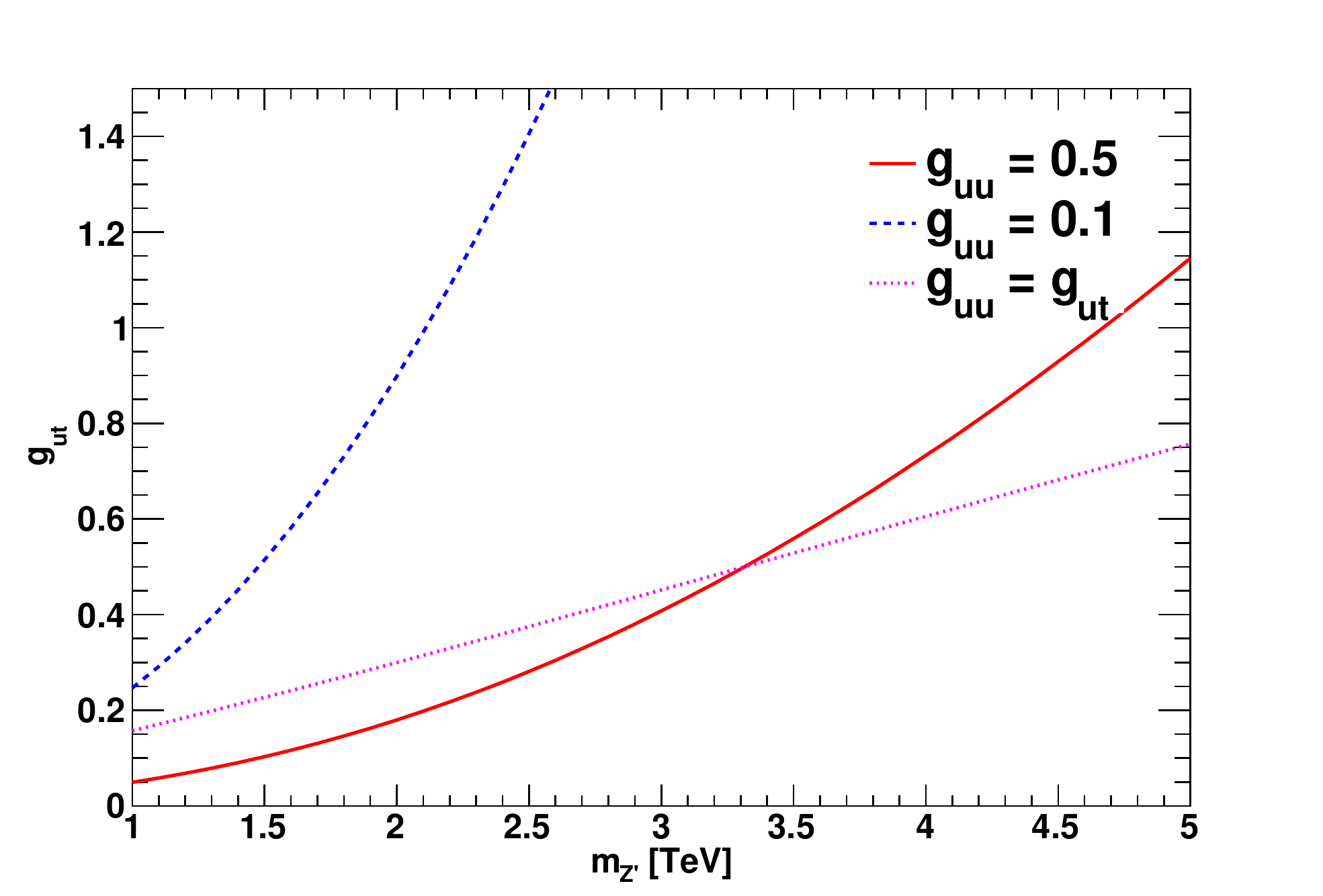} 
		\caption{ The contour plots of $g_{ut}$ vs $m_{Z'}$ for
			signal significance $S/\sqrt{B}=2$ at the production level
			 for  $\int{\mathscr L}dt=137$ fb$^{-1}$ at $\sqrt{s}=13$ TeV.
			The red, blue, and purple lines correspond to the cases 
			of $g_{uu}=0.5$, $g_{uu}=0.1$, and $g_{ut}=g_{uu}$, respectively.
		}
		\label{fig:gut-tj}
	\end{center}
\end{figure}

Note that the cross section for the same-sign top quark pair production is
proportional to $g_{ut}^4$, while that for the single top quark production
is to $g_{ut}^2 g_{uu}^2$. 
The cross section for the single top quark production strongly
depends on the value of $g_{uu}$ as well as $g_{ut}$.
For larger $g_{uu}$, the single top quark production may be more sensitive
to probe the top FCNCs, while the same-sign top quark pair production could be more sensitive
for relatively smaller $g_{uu}$.

To calculate the sensitivity, we need to calculate the SM backgrounds
for the single top quark production. 
We take into account $pp\to tj$, $\bar{t}j$, $tjj$, $\bar{t}jj$, $t\bar{b}$,
and $\bar{t}b$ processes with the cuts on jets: $p_T \ge 30$ GeV and $|\eta|<2.7$.
The bounds on the couplings $g_{uu}$ and $g_{ut}$ are determined by the ratio of signal to SM backgrounds, $S/\sqrt{B}$, 
at $2\sigma$ level. The $b$-jet identification efficiency  is assumed to be 50\,\%.

In Fig.~\ref{fig:gut-tj}, we depict the contour plots of $g_{ut}$ vs $m_{Z'}$ for
signal significance $S/\sqrt{B}=2$ at the production level
for $\int{\mathscr L}dt=137$ fb$^{-1}$ at $\sqrt{s}=13$ TeV.
The red, blue, and purple lines correspond to the cases 
of $g_{uu}=0.5$, $g_{uu}=0.1$, and $g_{ut}=g_{uu}$, respectively.

Here, we compare the bounds from the single top quark pair production with those
from the same-sign top quark pair production.
As shown in Fig.~\ref{fig:ss}, the bound on $g_{ut}$ could reach about $0.046$ for $m_{Z'}=2$ TeV with the integrated luminosity 
$\int {\mathscr L} dt=
137$ fb$^{-1}$ in the same-sign top quark pair production.
For $m_{Z'}=5$ TeV, the bounds on $g_{ut}$ could be about $0.10$,
while the bounds on $g_{ut}$ from the single top quark production
depend on  $g_{uu}$ as we already mentioned.
For $g_{uu}=0.5$ and $m_{Z'}=2$ TeV, the bounds on $g_{ut}$ can reach $0.18$ with the integrated luminosity $\int {\mathscr L} dt=
137$ fb$^{-1}$ in the single top quark production.
However, for $g_{uu}=0.1$ and $m_{Z'}=2$ TeV, the bounds on $g_{ut}$
could reach $0.90$, which is about five times larger than the $g_{uu}=0.5$ case. To obtain bounds similar to those
in the same-sign top quark pair production, 
$g_{uu}$ must be about $2.0$, but this value is ruled out
by the dijet production at the LHC, as we will show soon.
Therefore, we conclude that the same-sign top quark pair production is
more capable of yielding more stringent bounds on the top FCNC coupling
than the single top quark production.

\begin{figure}[tb]
	\begin{center}
		\includegraphics[width=75mm]{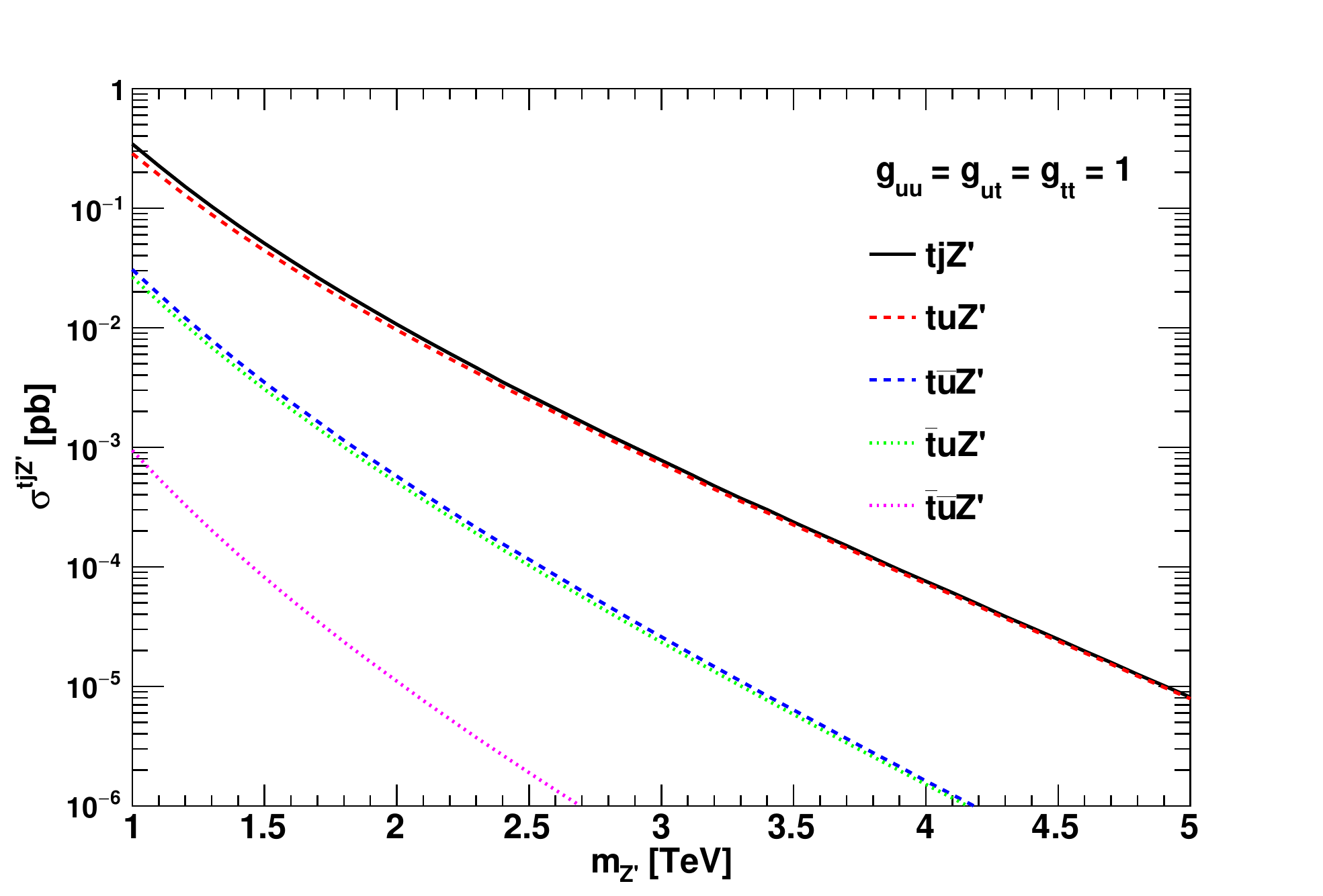} 
		\includegraphics[width=75mm]{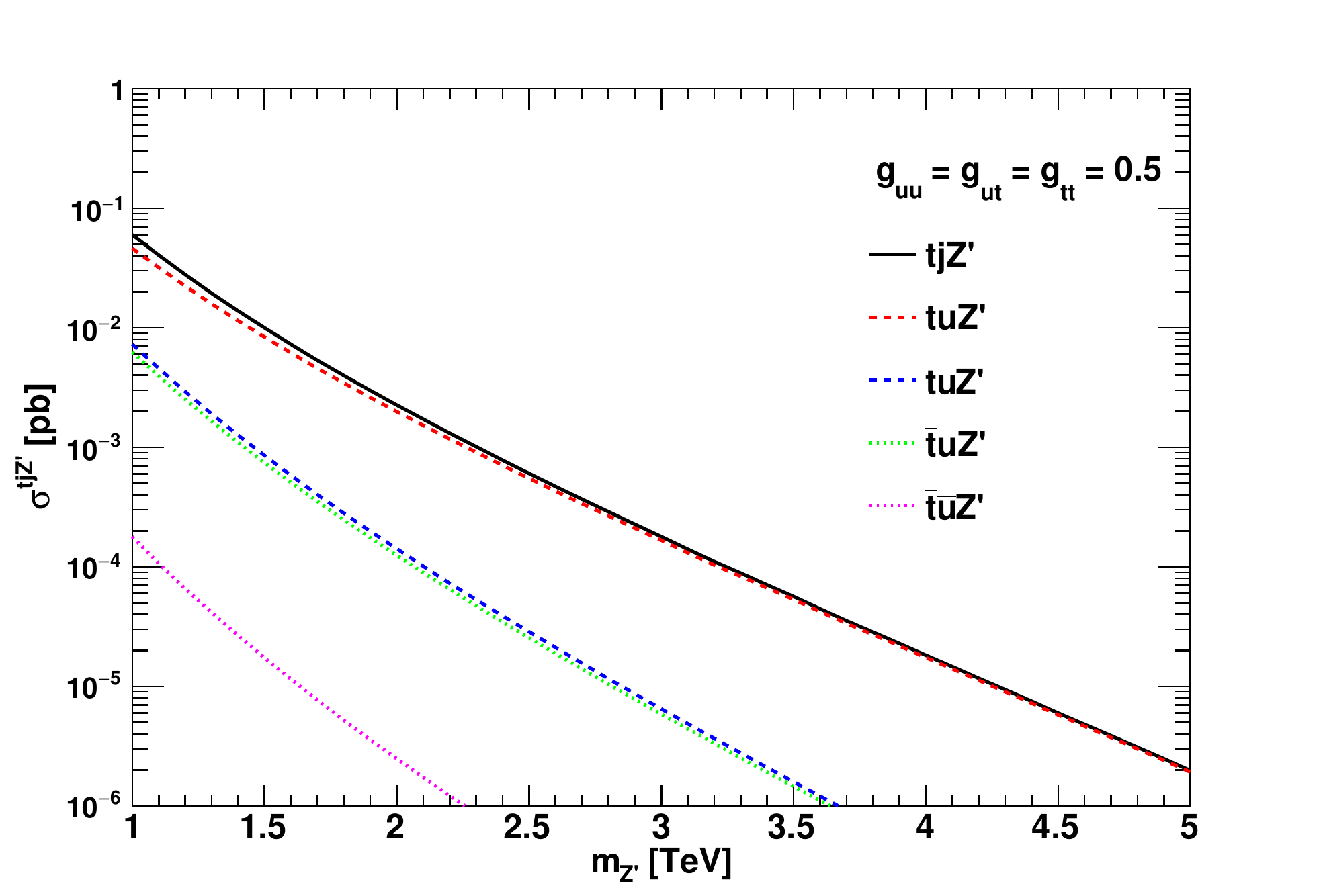} 
		\caption{ The cross sections for $pp\to t j Z'$ (in pb) for 
			$g_{ij}=1$ (left) and $g_{ij}=0.5$ (right) as a function of $m_{Z'}$ (in TeV) at $\sqrt{s}=13$ TeV.
			The red, blue, green, and purple lines correspond to $tu Z'$, $t\bar{u}Z'$, $\bar{t}u Z'$, and $\bar{t}\bar{u}Z'$ production, respectively, and the black
			line is the sum of them.	}
		\label{fig:tjzp}
	\end{center}
\end{figure}

\subsection{Radiative $Z'$ production}
\label{radiative}%

In this subsection, we take into account the $Z'$ production radiated from
the $u$ or $t$ quark line, where relevant processes are $pp\to t \bar{u} Z'$
or $\bar{t} u Z'$~\cite{Gupta:2010wt,Hou:2017ozb}. 
These processes contain the FCNC coupling $g_{ut}$. 
The emitted $Z'$ boson decays into 
$u\bar{u}$ ($t\bar{t}$, $u\bar{t}$, $t\bar{u}$),
a scalar and a gauge boson, or two gauge bosons.

Before the discussion on the $Z'$ production,
we briefly compare the radiative $Z'$ production in this work with the radiative Higgs ($H$) production by the top-Higgs FCNC through $pp \to t\bar{u} H$ or $\bar{t}u H$~\cite{Atwood:2013ica}.
The latter is dominated by the on-shell production of a $t\bar{t}$ pair,
followed by $t(\bar{t})\to H u(\bar{u})$ decays.
The cross section is enhanced by on-shell property of the intermediate $t$ or $\bar{t}$
because the Higgs boson is lighter than the top quark.
The Higgs boson is measured through its decays into $b\bar{b}$, $WW^*$, or $ZZ^*$ pairs,
which are governed by the SM interactions. Therefore, the total cross section for the radiative $H$ production is
proportional to $y_{tu}^2$, where $y_{tu}$ is the Yukawa coupling
responsible for the top-Higgs FCNC. 
In comparison with the radiative $H$ production, 
the same-sign top quark pair production proceeds through a $t$-channel exchange of
a Higgs boson in the parton process $uu\to tt$ or $\bar{u}\bar{u}\to \bar{t}\bar{t}$.
Then, the cross section
for the same-sign top quark pair production becomes proportional to $y_{tu}^4$.
Because of the intermediate on-shell top quark and dependence on the top-Higgs FCNC coupling, the radiative Higgs production can provide more stringent bounds
on the top-Higgs FCNC than the same-sign top quark pair production~\cite{Atwood:2013ica}.

The previous mechanism does not work in the radiative $Z'$ production with a $t$-$u$-$Z'$ FCNC coupling and a $Z'$ boson heavier than the top quark. Here, the $Z'$ boson is radiated from off-shell $u(\bar{u})$ or 
$t(\bar{t})$ quarks so that
production amplitudes are suppressed at least by a factor of $O(m_t^2/\hat{s})$, where $\hat{s}$ is the center-of-momentum energy
of parton processes.
Another point which one must consider is that the produced $Z'$ boson 
decays into $u\bar{u}$, $t\bar{t}$, $u\bar{t}$, $t\bar{u}$, or a pair of bosons.
Then, the cross section for the radiative $Z'$ production followed by the decay
of the $Z'$ boson would be proportional to ${g'}^4$, which is of  the same order as that for the same-sign top quark pair production. Therefore, we could not expect a dramatic increase
of the sensitivity to the FCNC coupling in the radiative $Z'$ production.

The radiative $Z'$ production processes which we take into account are
$pp\to t \bar{t} (u\bar{u})$ followed by subsequent off-shell decays $t(\bar{t})\to Z' u(\bar{u})$ or $u(\bar{u})\to Z' t(\bar{t})$.
The subsequent decays contain a FCNC coupling $g_{ut}$, while $t\bar{t}$ or $u\bar{u}$ production amplitudes
have a dependence on $Z'$-involved couplings $g_{ij}^2$,
where $i,j=u$ or $t$, or do not contain $Z'$ interactions.
Therefore, the total cross section can be expressed as the sum of terms
proportional to $g_{ut}^2$, $g_{ut}^2 g_{ij}^2$, and $g_{ut}^2 g_{ij}^4$.
In Fig.~\ref{fig:tjzp}, we show the cross sections for the radiative $Z'$ production,
$pp\to t j Z'$ (in pb) for $g_{ij}=1$ (left) and $g_{ij}=0.5$ (right) as functions of $m_{Z'}$ (in TeV) at $\sqrt{s}=13$ TeV.
The red, blue, green, and purple lines correspond to $tuZ'$, $t\bar{u}Z'$, 
$\bar{t}uZ'$, and $\bar{t}\bar{u}Z'$, respectively, and the black line
is the sum of them.
As was discussed in the single top quark production, the cross section for the $tuZ'$ production is the largest  
 because of the parton density of the incident $u$ quark.
The small difference between the $t\bar{u}Z'$ and $\bar{t}uZ'$ processes stems from
the kinematic cuts ($p_T \ge 30$ GeV and $|\eta| \le 2.7$)
as was discussed in the previous section.
The sum of the cross sections for the radiative $Z'$ production is about
$0.3$ pb for $m_{Z'}=1$ TeV and $g_{ij}=1$, but  it decreases to $10^{-2}$ fb for $m_{Z'}=5$ TeV.

Since the cross sections for the radiative $Z'$ production are given 
in terms of $g_{ut}^2$, $g_{ut}^4 g_{ij}^2$, $g_{ut}^2 g_{ij}^4$, and $g_{ut}^6$,
we cannot express them in a simple form like the same-sign top quark pair
production or single top quark production.
However, we find that the contribution of the $g_{ut}^2$ terms to the total cross section is dominant over the other terms.
For $m_{Z'}\ge 2$ TeV, its contribution is more than $85$\,\% and
even for a lighter $Z'$ with $m_{Z'}=1$ TeV, it becomes about $71$\,\%.
Hence, for $m_{Z'}=2$ TeV, the cross sections can be
approximated as 
\begin{subequations}
\begin{eqnarray}
\sigma(pp\to t u Z') &\sim& 9.6 \, g_{ut}^2~\textrm{fb},
\\
\sigma(pp\to t \bar{u} Z') &\sim& 0.57\, g_{ut}^2~\textrm{fb},
\\
\sigma(pp\to \bar{t} u Z') &\sim& 0.51\, g_{ut}^2~\textrm{fb},
\\
\sigma(pp\to \bar{t} \bar{u} Z') &\sim& 0.011 \,g_{ut}^2~\textrm{fb},
\end{eqnarray}
\end{subequations}
where the sum of the cross sections is 
\begin{equation}
\sigma(pp\to \bar{t} j Z') \sim 10.7\, g_{ut}^2~\textrm{fb}.
\end{equation}
If one considers the decay of the $Z'$ boson, for example, $Z'\to u\bar{u}$,
the cross section would be $O(1)$ fb for $g_{ut}=1$.
The dominant SM backgrounds for the radiative $Z'$ production followed by
the $Z'$ decay into $u\bar{u}$ are $pp\to t (\bar{t})jjj$ and $pp\to t \bar{b}(\bar{t}b)jj$, where the $b$-jet identification efficiency must
be taken into account in the latter case. 
Without implementing any cut on the invariant mass of a pair of final jets,
the cross sections of the SM backgrounds reach $O(100)$ pb.
If one implements the condition that the invariant mass of a pair of jets
is between $m_{Z'}-\Gamma_{Z'}/2$ and $m_{Z'}+\Gamma_{Z'}/2$,
the SM backgrounds could become negligible. The situation for the SM backgrounds with the cuts on the invariant mass of  a pair of jets
is similar to that for the same-sign top quark pair production.
If we compare the radiative $Z'$ production with the same-sign top quark pair
production, the latter has a larger cross section by a factor of $O(10^2)$.
Therefore, we conclude that the same-sign top quark pair production is
more capable of yielding better bounds on the FCNC coupling $g_{ut}$.

\begin{figure}[tb]
	\begin{center}
		\includegraphics[width=75mm]{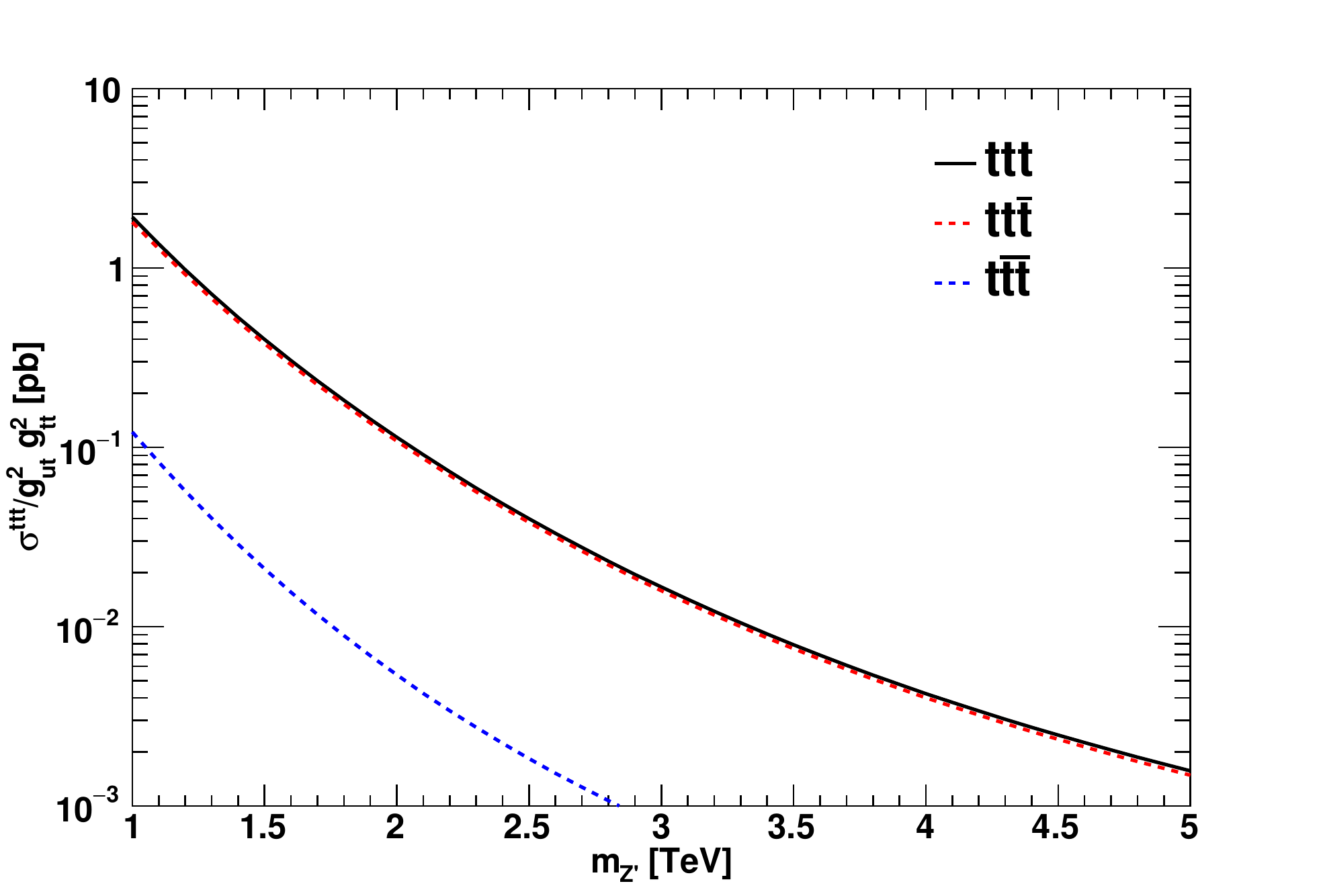} 
		\caption{ The cross sections for the triple top quark production (in pb) divided by $g_{ut}^2 g_{tt}^2$ as a function of $m_{Z'}$ (in TeV) at $\sqrt{s}=13$ TeV.
			The red and blue lines correspond to $tt\bar{t}$ and $t\bar{t}\bar{t}$ production, respectively, and the black
			line is the sum of them.	}
		\label{fig:ttt}
	\end{center}
\end{figure}

\subsection{Triple top quark production}

Another possible test for the top FCNCs is the triple top quark production 
via $u g \to t Z' \to t t \bar{t}$ or $\bar{u}g \to \bar{t} Z' \to t\bar{t}\bar{t}$. The production amplitudes are proportional to $g_{ut} g_{tt}$, where
$g_{tt}$ can be constrained from the top quark pair production and/or
four top quark production. In the case of $g_{tt} \ll g_{uu, ut}$, 
the cross section for the triple top quark production would be negligible in comparison with the single top quark production
whose production amplitude is proportional to $g_{uu}g_{ut}$.
On the other hand, in the case of $g_{uu} \ll g_{tt, ut}$,
the single top quark production would be less sensitive to probe top FCNCs
than the triplet top quark production.
Therefore, the sensitivity of each process to the top FCNCs
strongly depends on the couplings $g_{uu}$ and $g_{tt}$.

For $m_{Z'}=2$ TeV, the cross sections for the triple top quark production at $\sqrt{s}=13$ TeV are obtained as
\begin{eqnarray}
\sigma(pp\to tt\bar{t}) &=& 0.11 \, g_{ut}^2 g_{tt}^2~\textrm{pb},\\
\sigma(pp\to t\bar{t}\bar{t}) &=& 0.0054 \, g_{ut}^2 g_{tt}^2~\textrm{pb},
\end{eqnarray}
where the latter is much suppressed because of the $\bar{u}$-quark parton density in a proton in comparison with that of the $u$ quark.
For $g_{ut} \sim g_{tt} \sim 0.5$, the sum of the triple top quark production cross sections
is about $7.1$ fb.

In Fig.~\ref{fig:ttt}, we plot the cross sections for the triple top quark
production (in pb) divided by $g_{ut}^2 g_{tt}^2$
as a function of $m_{Z'}$ (in TeV) at $\sqrt{s}=13$ TeV.
The red and blue lines correspond to $tt\bar{t}$ and $t\bar{t}\bar{t}$
production, respectively, and the black line is the sum of them.
For $m_{Z'}=1$ TeV, the cross section reaches about $O(1)$ pb, but
it becomes $O(1)$ fb for $m_{Z'}=5$ TeV.

\begin{figure}[tb]
	\begin{center}
		\includegraphics[width=75mm]{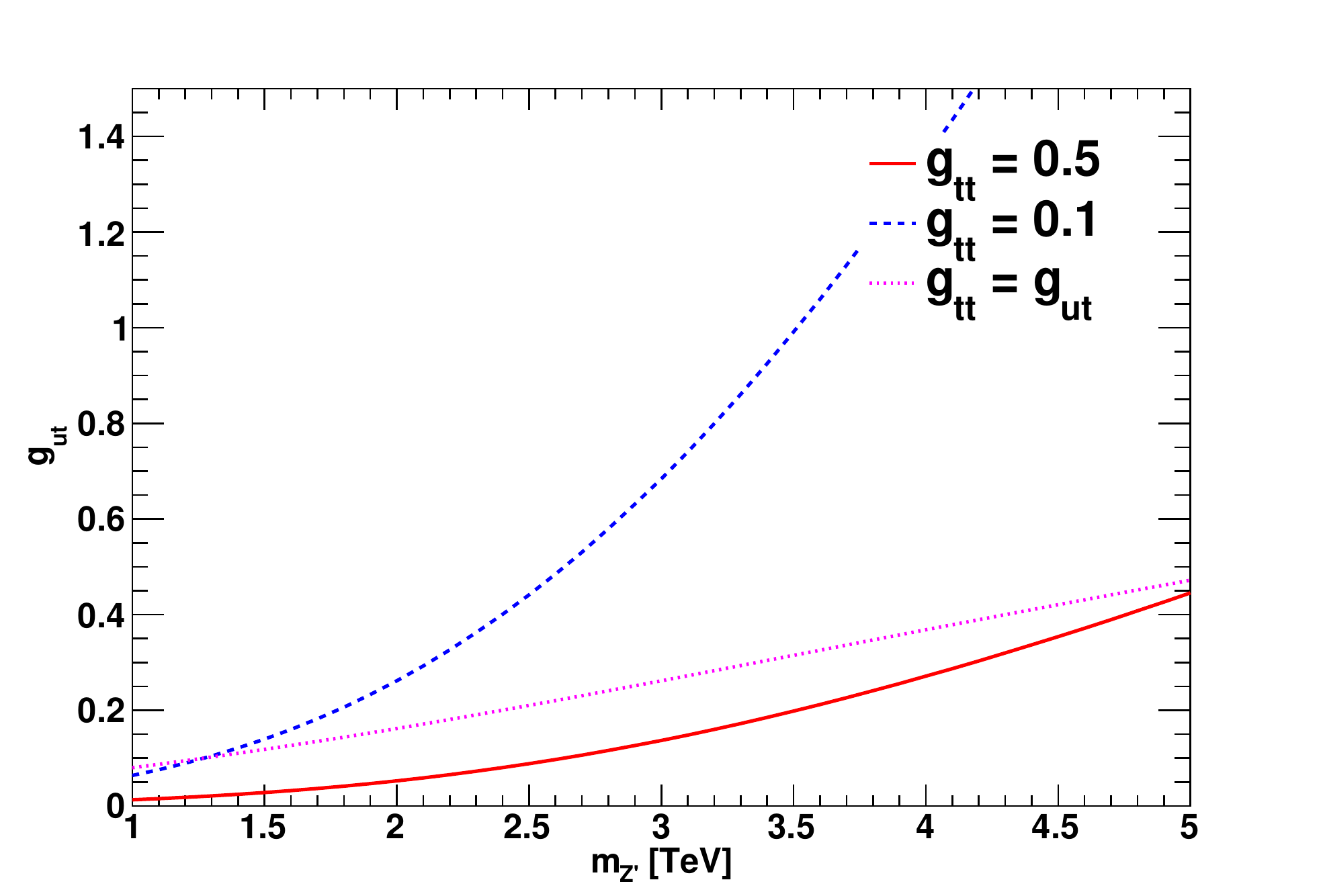} 
		\caption{ The contour plots of $g_{ut}$ vs $m_{Z'}$ for
			signal significance $S/\sqrt{B}=2$
			for  $\int{\mathscr L}dt=137$ fb$^{-1}$  at $\sqrt{s}=13$ TeV.
			The red, blue, and purple lines correspond to the cases 
			of $g_{tt}=0.5$, $g_{tt}=0.1$, and $g_{tt}=g_{ut}$, respectively.
		}
		\label{fig:gut-ttt}
	\end{center}
\end{figure}

The SM backgrounds for the triple top quark production which we consider are
$pp\to  tt\bar{t}j$, $t\bar{t}\bar{t}j$, $tt\bar{t}b$, $t\bar{t}\bar{t}b$,
$tt\bar{t}\bar{b}$, and $t\bar{t}\bar{t}\bar{b}$, where one must consider
the $b$-jet identification efficiency for the last four processes.
With the cuts on jets, $p_T \ge 30 $ GeV and $|\eta|<2.7$, we calculate 
the cross sections of the SM backgrounds and the ratio of signal to 
backgrounds at $2\sigma$ level.
In Fig.~\ref{fig:gut-ttt}, we depict the contour plots of $g_{ut}$ vs $m_{Z'}$ for
signal significance $S/\sqrt{B}=2$
for the integrated luminosity $\int{\mathscr L}dt=137$ fb$^{-1}$ at $\sqrt{s}=13$ TeV.
The red, blue, and purple lines correspond to the cases 
of $g_{tt}=0.5$, $g_{tt}=0.1$, and $g_{ut}=g_{tt}$, respectively.

Here, we compare the bound on $g_{ut}$ from the triple top quark  production with the one
from the same-sign top quark pair production.
The bound on $g_{ut}$ from the triple top quark production
depends on the value of $g_{tt}$ as we already mentioned.
For $g_{tt}=0.5$ and $m_{Z'}=2$ TeV, the bound on $g_{ut}$ can reach $0.055$ with the integrated luminosity $\int {\mathscr L} dt=137$ fb$^{-1}$ in the triple top quark production, while
for $g_{tt}=0.1$ and $m_{Z'}=2$ TeV, the bound on $g_{ut}$ 
reaches $0.27$. To obtain bounds comparable to those
in the same-sign top quark pair production, 
$g_{tt}$ should be about $0.68$.
Therefore, we conclude that the same-sign top quark pair production tends to yield more stringent bounds on the top FCNC coupling
than the triple top quark production for $g_{tt}\lesssim 0.68$.
However, for $g_{tt}\gtrsim 0.68$, the triple top quark production
might provide more stringent bounds on $g_{ut}$.
We note that this comparison is based on the numerical analysis
at the parton level, where the $t$ or $\bar{t}$ quarks are
completely reconstructed from the data. In the analysis at detector level, it would be difficult to reconstruct all the top quarks, in particular,
in the multitop production because of the missing energy for the semileptonically decaying top quarks
and combinatoric problems of jets of hadronically decaying top quarks. 
We will perform the numerical analysis at detector level in Sec.~\ref{sec;simulation}, and it will turn out that the bounds on the couplings in the triple top quark production could be much larger than the results in this subsection.

\begin{figure}[tb]
	\begin{center}
		\includegraphics[width=75mm]{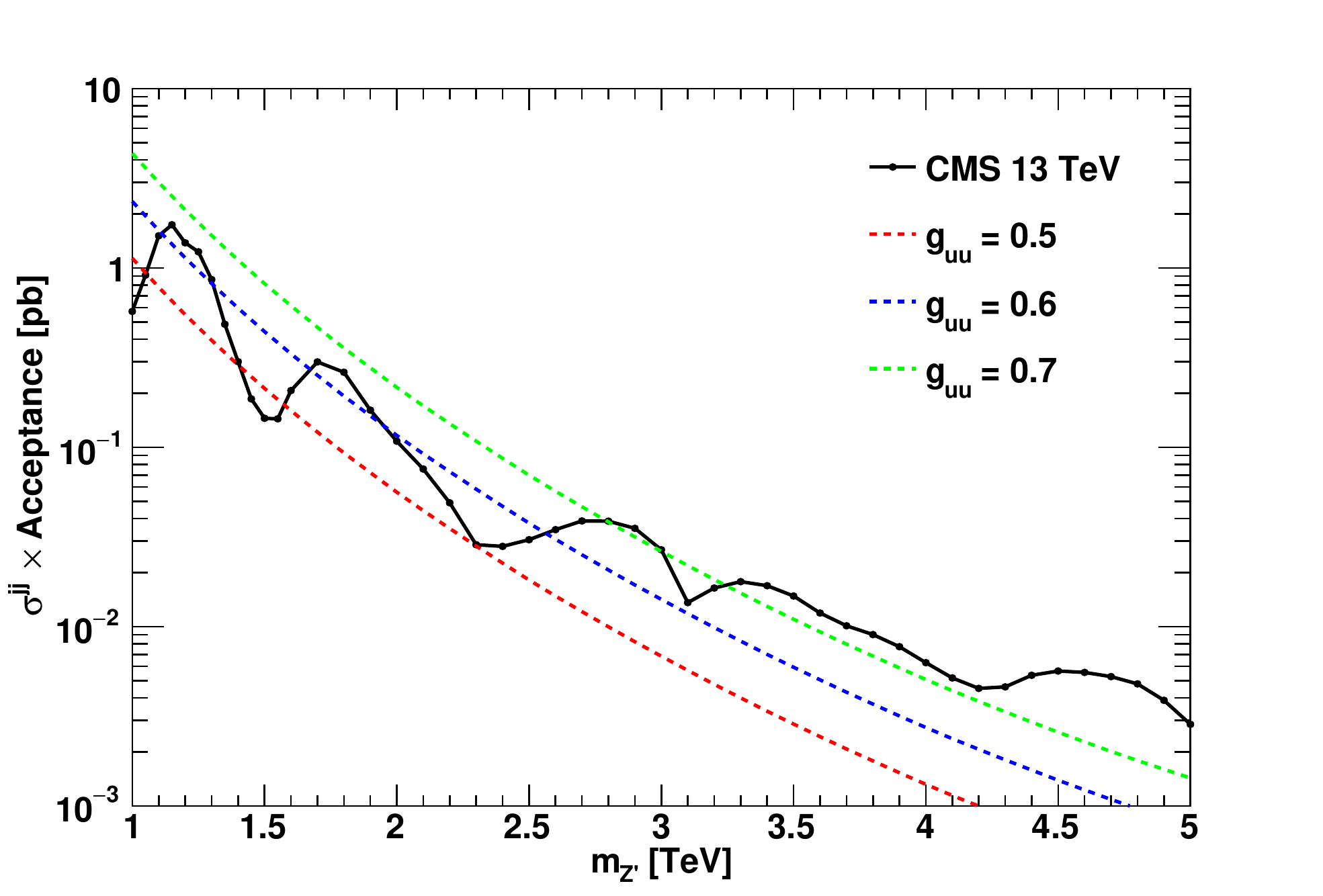} 
		\caption{ The cross sections for the dijet production (in pb) via a $Z'$ exchange as a function of $m_{Z'}$ (in TeV) at $\sqrt{s}=13$ TeV.
			The red, blue, and green lines correspond to $g_{uu}=0.5$, $0.6$, and $0.7$,
			respectively. The black line is the observed 95\,\% CL upper limits at CMS~\cite{Sirunyan:2018xlo}.}
		\label{fig:dijet}
	\end{center}
\end{figure}

\subsection{Dijet production}

The dijet production is one of the best channels to probe an $s$-channel resonance
like, in particular, a leptophobic $Z'$ boson.
In this work, the $Z'$ boson is taken into account as a leptophobic $Z'$ boson 
by assuming that mixing between the SM neutral gauge bosons and the $Z'$ boson is negligible.
The dijet production through the $Z'$ resonance proceeds in the parton process
$u \bar{u} \to Z' \to u\bar{u}$, whose amplitude is proportional to $g_{uu}^2$.
Then, the cross section for the dijet production through the $Z'$ resonance 
for $m_{Z'}=2$ TeV at $\sqrt{s}=13$ TeV is calculated as
\begin{equation}
\sigma(pp\to Z'\to jj)=0.92\, g_{uu}^4~\textrm{pb},
\end{equation}
where the $K$ factor is taken to be $1.3$~\cite{Sirunyan:2018xlo} and the cuts on jets are implemented.

In Fig.~\ref{fig:dijet}, we depict the cross sections for the dijet production
at $\sqrt{s}=13$ TeV.
The red, blue, and green lines correspond to
$g_{uu}=0.5$, $0.6$, and $0.7$, respectively.
The black line is the observed 95\,\% CL upper limits with the integrated luminosity, $\int {\mathscr L}=36$ fb$^{-1}$, at CMS~\cite{Sirunyan:2018xlo}.
From the CMS data, the region $g_{uu}\lesssim 0.5$ is allowed for $m_{Z'}\ge 1$ TeV. For $m_{Z'}\gtrsim 3.3$ TeV, $g_{uu}\lesssim 0.7$ is allowed.
With more accumulated data, the upper bounds could be much improved.
However, it might not be capable of probing the $Z'$ boson with 
the small coupling $g_{uu}\ll 1$ even with more data.
Then, the search for the $Z'$ boson through the top FCNC processes,
for example, the same top quark pair production, may be a more probable
channel to probe the $Z'$ boson in such a case.

\begin{figure}[tb]
	\begin{center}
		\includegraphics[width=75mm]{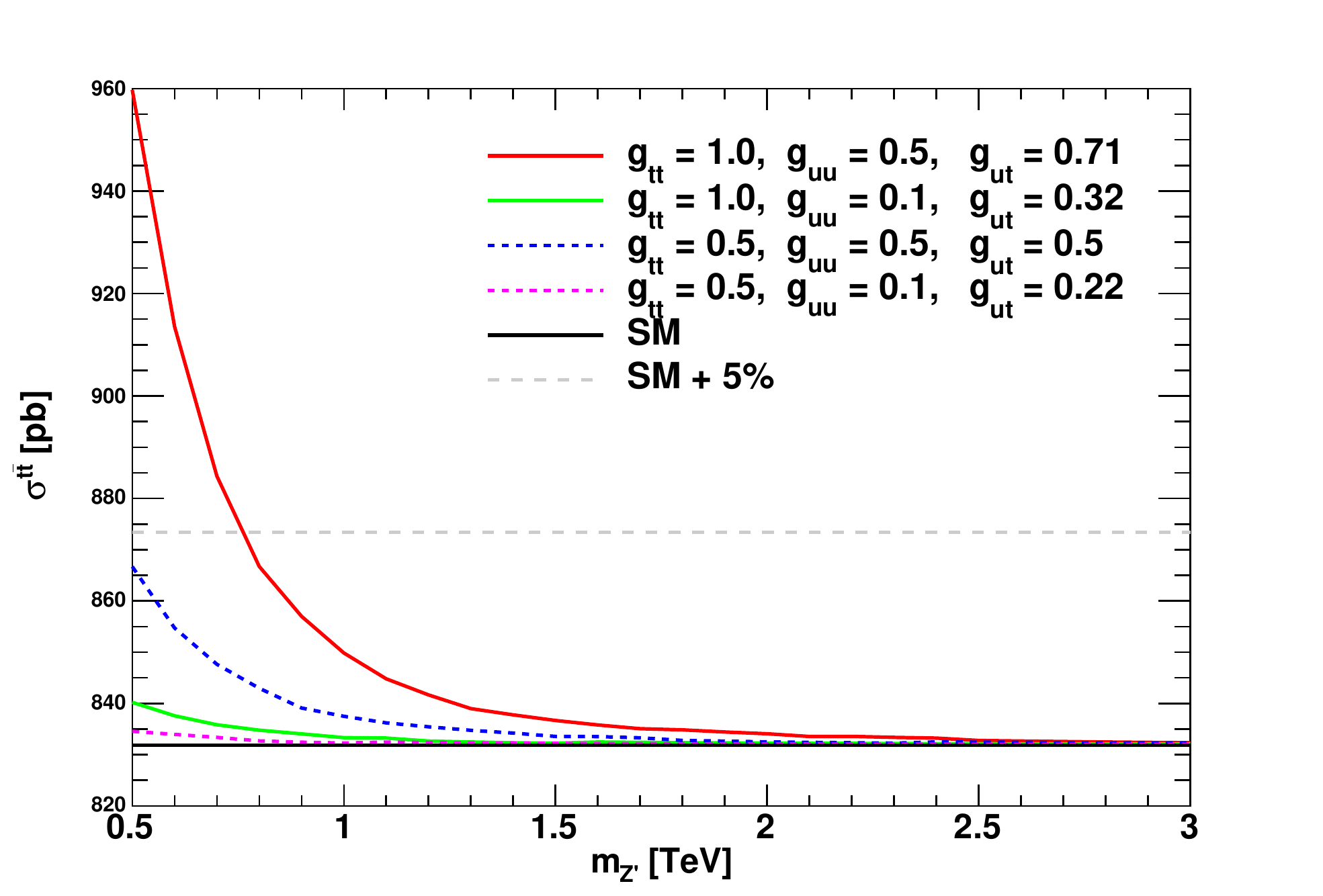} 
		\includegraphics[width=75mm]{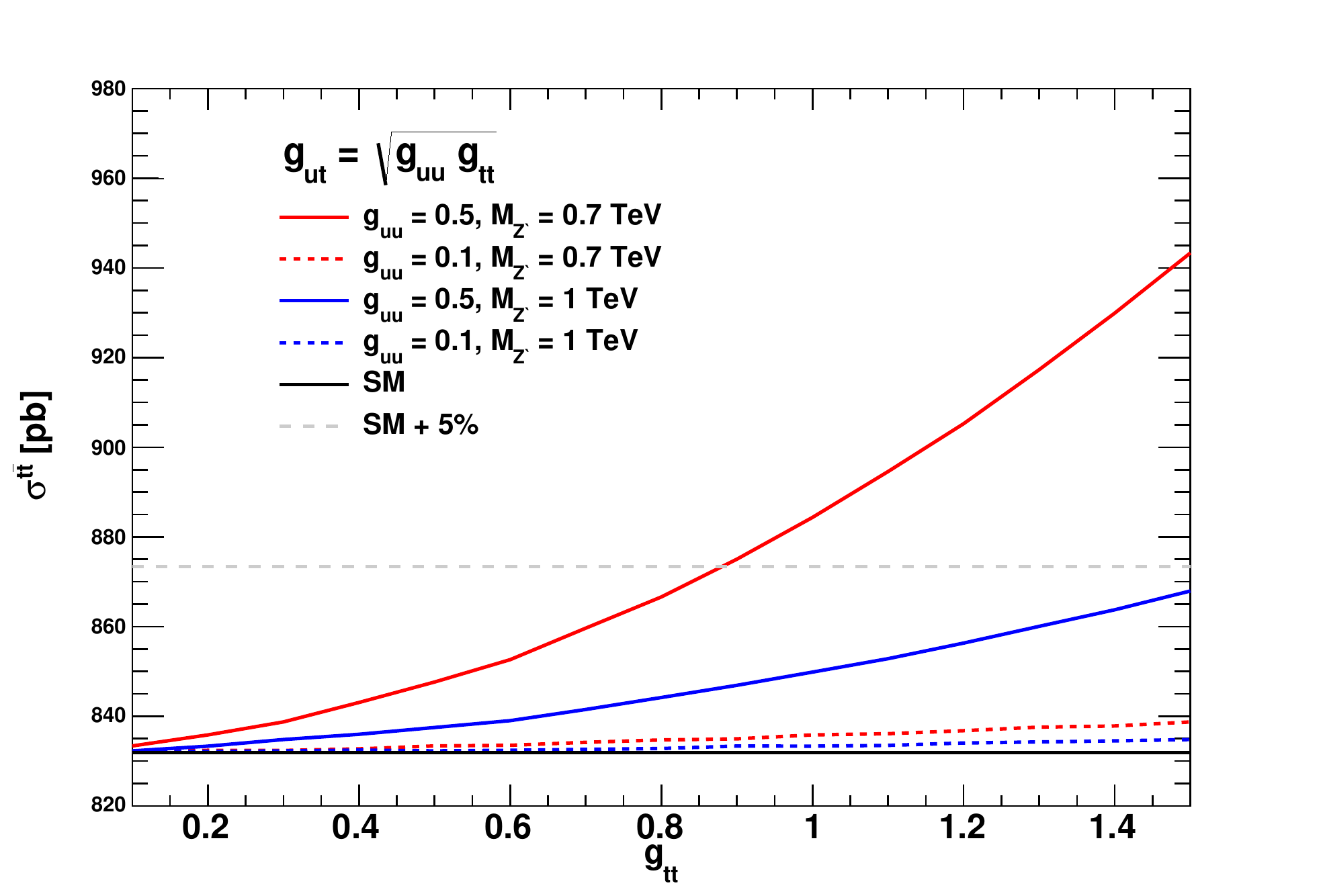} 
		\caption{ The cross sections for the top quark pair production
			(in pb) as functions of $m_{Z'}$ in TeV (left)
			and $g_{tt}$ (right) at $\sqrt{s}=13$ TeV.
			The black line is the SM prediction at leading order.
			}
		\label{fig:ttbar}
	\end{center}
\end{figure}

\subsection{Top quark pair production}

In this subsection, we consider the top quark pair production  $pp\to t\bar{t}$.
In principle, one can consider the top quark pair production via
a $Z'$ resonance like the dijet production in the previous section.
So far at the c.m. energy $\sqrt{s}=13$ TeV, there are no results 
for a resonance search in the top quark pair production except 
for the heavy Higgs boson search 
ranging from $400$ to $750$ GeV in CMS experiments~\cite{Sirunyan:2019wph}.
Thus, in this subsection, we concentrate on the effects of the $Z'$ boson
on the total cross section for the top quark pair production.

The parton process relevant to the top quark pair production
is $u\bar{u}\to t\bar{t}$, where the $Z'$ boson can be exchanged
in an $s$-channel as well as in a $t$-channel.
Then, three couplings $g_{uu}$, $g_{ut}$, and $g_{tt}$ are involved
in the top quark pair production.
The $s$ channel exchange amplitude is proportional to $g_{uu}g_{tt}$,
while the $t$ channel one is to $g_{ut}^2$.
In Fig.~\ref{fig:ttbar}, we depict the total cross sections
for the top quark pair production (in pb) as functions of
$m_{Z'}$ in TeV (left) and $g_{tt}$ (right)
for various values of $g_{uu}$ at $\sqrt{s}=13$ TeV.
Here, the $K$ factor is taken to be $1.6$.
$g_{ut}$ is determined by $g_{ut}=\sqrt{g_{uu}g_{tt}}$, but we note that
this relation is valid for the mixing between two right-handed quarks 
$u_R$ and $t_R$.
If one considers the mixing among three right-handed up-type quarks,
$g_{ut}$ could be set to be a free parameter.
It should be noted that the $Z'$ boson does not contribute to the top quark pair production for $g_{uu}\sim g_{ut} \sim 0$.

In Fig.~\ref{fig:ttbar}, the horizontal black line is the SM prediction
for the top quark pair production. 
In experiments, statistical uncertainties in the top quark pair production are well controlled and
much smaller than the other ones, systematic and luminosity uncertainties~\cite{top2018summary}. 
It seems that the sum of the systematic and luminosity uncertainties in quadrature is about $5$\,\%,
regardless of the decay channels of the top quark~\cite{top2018summary}, which is at least ten times larger than statistical uncertainties in the lepton+jet channel
of the top decay. Thus we do not perform the analysis
for the SM backgrounds in the top quark pair production.
Instead, we consider $5$ \% deviation from the SM prediction,
which is denoted by the gray line in Fig.~\ref{fig:ttbar}.
In the left panel of Fig.~\ref{fig:ttbar}, we plot the total cross sections for several
values of $g_{ij}$. For $g_{uu}=g_{tt}=0.5$, the total cross section cannot reach
the $5$\,\% uncertainty line for  $m_{Z'} \gtrsim 500$ GeV.
In the right panel of Fig.~\ref{fig:ttbar}, the total cross sections are depicted
as a function of $g_{tt}$ for $m_{Z'}=700$ GeV (red) and $1$ TeV (blue), respectively.
For $m_{Z'}=700$ GeV and $g_{uu}=0.5$, the cross section can reach the gray line 
for $g_{tt} \sim 0.9$. Therefore, we conclude that 
the light $Z'$ boson and large $g_{tt}$ couplings are required for the $Z'$ boson 
to be probed in the top quark pair production.
It is worthwhile to mention that if $g_{uu}$ is negligible it might be difficult to probe the $Z'$ boson in the top quark pair production.

\begin{figure}[tb]
	\begin{center}
		\includegraphics[width=75mm]{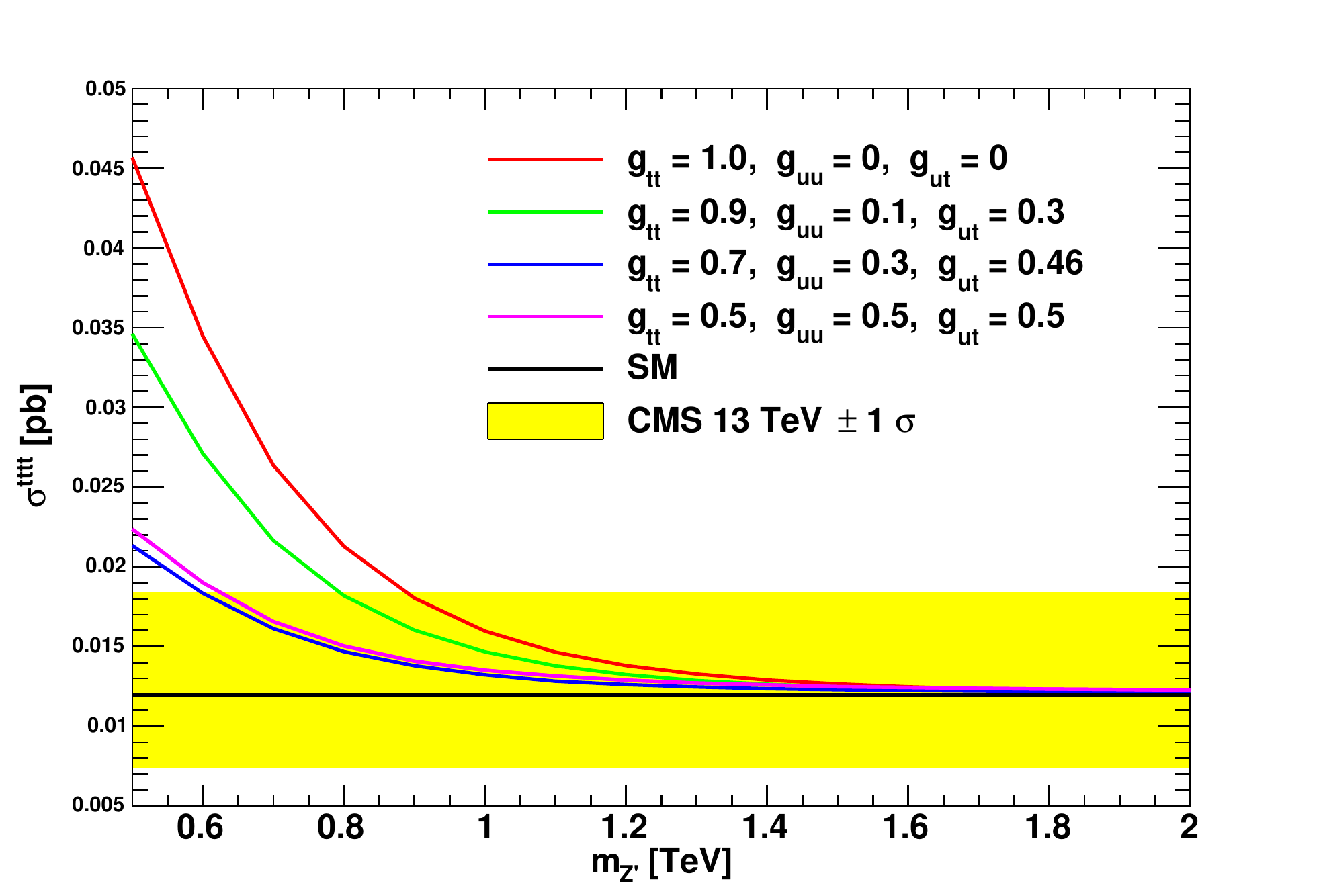} 
		\includegraphics[width=75mm]{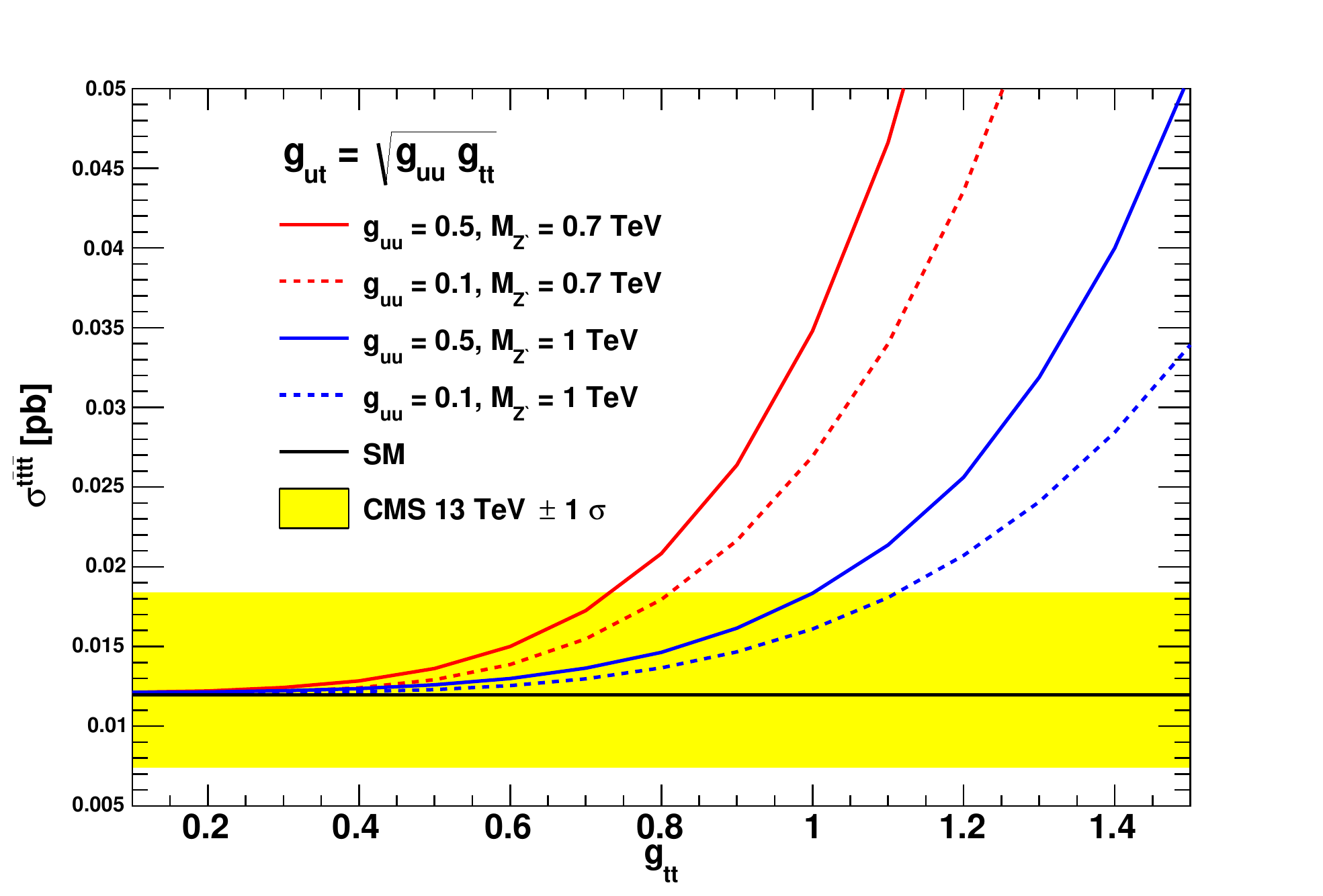} 
		\caption{ The cross sections for the four top quark production
			(in pb) as functions of $m_{Z'}$ in TeV (left)
			and $g_{tt}$ (right) at $\sqrt{s}=13$ TeV.
			The yellow band is the measured value at CMS within 95\,\% CL~\cite{Sirunyan:2019wxt},
			and the black line is the SM prediction.
		}
		\label{fig:4t}
	\end{center}
\end{figure}

\subsection{Four top quark production}
The last process which we consider in this work is the four top quark production
at the LHC. 
The $Z'$ boson can affect the four top quark production 
through a $Z'$ exchange between top quark pairs or between the incident $u\bar{u}$ pair and produced $t\bar{t}$ pair.
Then, all three couplings, $g_{uu}$, $g_{ut}$, and $g_{tt}$, are involved
in the four top quark production.
Since the amplitudes may involve one or two $Z'$ propagators,
the total cross section can be expressed in terms of $g_{ij}^k$ ($k=0,2,4,6,8$)
if one also considers the SM interactions.
One of the merits of the four top quark production is that, unlike the top quark pair production, the $Z'$ boson can contribute to
the four top quark production even for $g_{uu}=g_{ut}=0$.
This can occur by the $t\bar{t}$ pair production, followed by radiation of a $Z'$ boson
from $t$ or $\bar{t}$ and its subsequent decay into another
$t\bar{t}$ pair. Therefore, one may expect the $Z'$ signal 
even though one cannot observe it in the top quark pair production
and same-sign top quark pair production, in which $g_{uu}$ and/or $g_{ut}$ are involved.
On the other hand, the production mechanism of this process is similar to the radiative $Z'$ production
discussed in Sec.~\ref{radiative}. Thus, one could not expect a large deviation
from the SM prediction for large $m_{Z'}$.

In Fig.~\ref{fig:4t}, we depict the total cross sections for the four top quark production
in pb as functions of $m_{Z'}$ in units of TeV (left) and $g_{tt}$ (right)
at $\sqrt{s}=13$ TeV. 
The black solid line is the SM prediction.
Here, we set the $K$ factor to be $1.3$~\cite{Frederix:2017wme}.
The yellow band is the measured value at CMS with the integrated luminosity $137$ fb$^{-1}$ within 95\% CL: $\sigma = 12.6^{+5.8}_{-5.2}$ fb~\cite{Sirunyan:2019wxt}, which is consistent with the SM prediction
$12.0^{+2.2}_{-2.5}$ fb~\cite{Frederix:2017wme}.
The total cross sections are depicted
for several combinations of $g_{uu}$ and $g_{tt}$, and in the first four cases,
$g_{uu}\le 0.5$ are chosen by considering the bound on $g_{uu}$
from the dijet production.
$g_{ut}$ is determined by $g_{ut}=\sqrt{g_{uu}g_{tt}}$. 
In the case of mixing of three right-handed up-type quarks, 
$g_{ut}$ could be chosen freely, as we have argued in Sec.~\ref{sec;setup}.

As shown in Fig.~\ref{fig:4t}, the cross section could be enhanced for
a small $Z'$ mass, for example, in the region of $m_{Z'} \lesssim 1$ TeV.
From the right panel of Fig.~\ref{fig:4t}, we also require a large $g_{tt}$ coupling 
to get a large deviation from the SM prediction.
As an example, $g_{tt}\gtrsim 0.7$ is required for $m_{Z'}=0.7$ TeV and $g_{uu}=0.5$. 
For a small $m_{Z'}$, the four top quark production may probe the $Z'$ boson.
However, for the $Z'$ mass larger than about $1.2$ TeV, this process would be difficult
to probe the $Z'$ boson, even though more data are accumulated at the LHC.

\section{Numerical analysis at detector level}
\label{sec;simulation}%

In this section, we perform numerical analyses at detector level
for the same-sign top qurk pair production and triple top quark production,
which might be the best candidate for the proof of the top FCNC coupling $g_{ut}$.
In the former case, the result of the detector simulation is not very different from that of the parton-level analysis
because the SM backgrounds are very small in comparison with generated signals.
However, for the triple top quark production, we find that the bound
for the coupling 
obtained in the previous section is much smaller than
that from the detector simulation in this section.
The main difference between two analyses comes from the estimation
of the SM backgrounds, which we will discuss later.

In the same-sign top quark pair production, 
we consider the process $pp\to tt(\bar{t}\bar{t})$
as well as $pp \to tt\bar{u} (\bar{t}\bar{t}u)$,
where the latter occurs from a $ug (\bar{u}g)$ collision.
For leptonical decays, $t\rightarrow bl^+\nu_l$ and 
$\bar{t} \rightarrow \bar{b} l^- \bar{\nu}_l$ ($l=e, \mu$),
the final signals contain two same-sign leptons and two $b$ jets
with missing transverse energy.
We produce the signal events
by making use of MADGRAPH~\cite{Alwall:2011uj,Alwall:2014hca}, Pythia~\cite{Pythia}, and Delphes~\cite{Delphes} for parton-level event generation, parton shower, hadronization, and detector simulation at the 13 TeV LHC.
We follow the analysis methods for the search for new physics signals with the signature of a same-sign lepton pair and $b$ jet by the ATLAS group~\cite{Aaboud:2018xpj} and require following conditions:
	
\begin{enumerate}[(i)]
\item exactly one same-sign lepton pair with each lepton $p_T$ larger than $28$ GeV,
\item at least one $b$-tagged jet with the $b$-jet $p_T$ larger than $25$ GeV,
\item missing transverse energy ($\slashed{E}_{T}$) lager than $40$ GeV,
\item $H_T$ larger than $750$ GeV,
\item the azimuthal angle separation between two leptons larger than $2.5$.
\end{enumerate} 
	
The $H_T$ is the scalar sum of the transverse momentum of all jets in an event. 
We find that about $0.3\%$ of the signal events passes the cuts listed above
and the ATLAS Collaboration shows that $23.9$ events of the SM background
survive after the selection for
$\int{\mathscr L}dt=36.1$ fb$^{-1}$.
We simply rescale the expected number of SM background events
$n_b=23.9\times (\int{\mathscr L}dt/36.1)$
for other integrated luminosities $\int{\mathscr L}dt$~\cite{Cao:2019qrb}.
	
\begin{figure}[tb]
\begin{center}
\includegraphics[width=75mm]{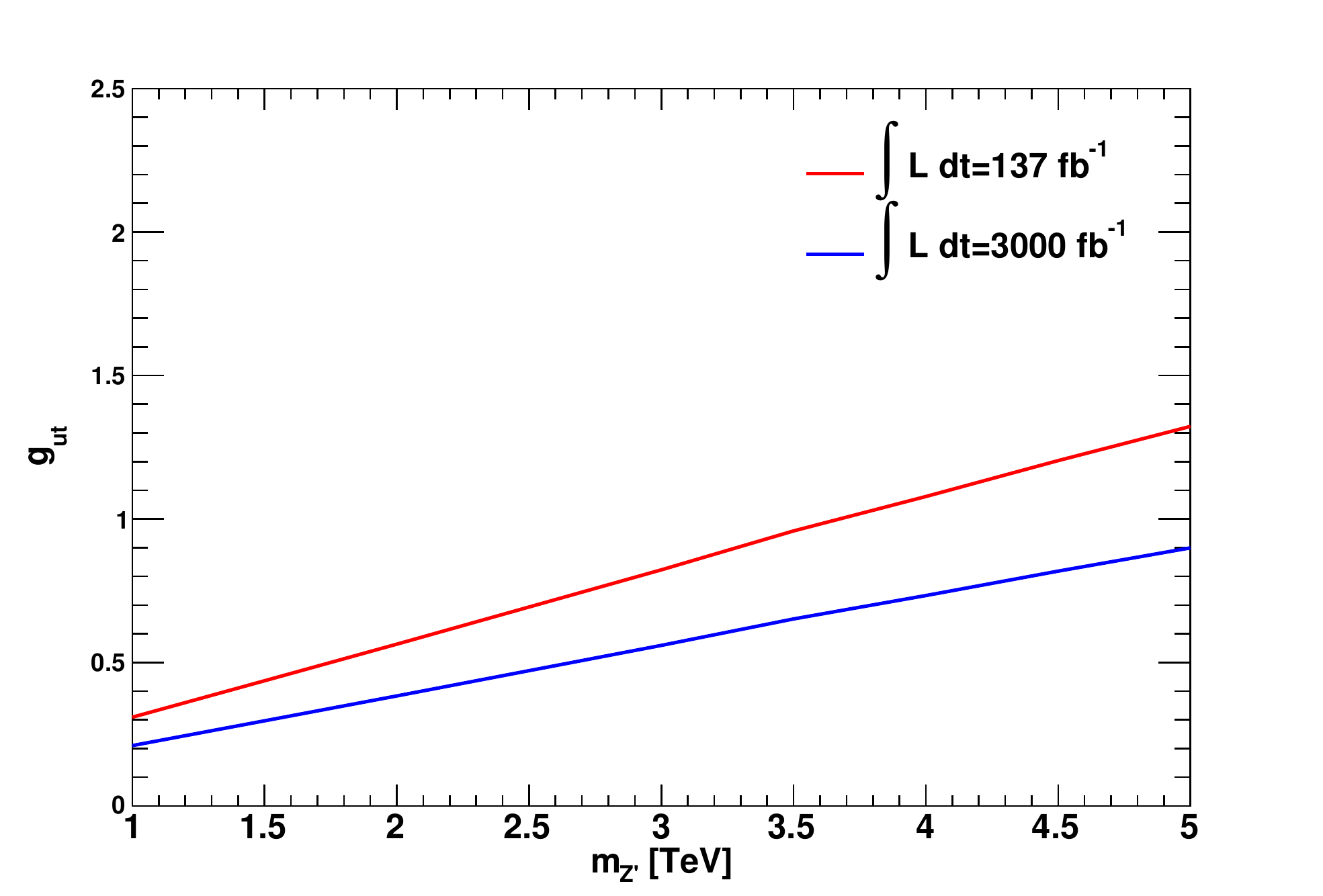} 
\caption{The contour plots of $g_{ut}$ vs $m_{Z'}$ (in TeV) for
signal significance $S/\sqrt{B}=2$ at the detector level (ATLAS)
for $\int{\mathscr L}dt=137$ fb$^{-1}$ (red) and $\int{\mathscr L}dt=3000$ fb$^{-1}$ (blue) at $\sqrt{s}=13$ TeV.
}
\label{fig:sstopATLAS}
\end{center}
\end{figure}
	
In Fig.~\ref{fig:sstopATLAS}, we depict the contour plots of $g_{ut}$ vs $m_{Z'}$ (in TeV) for
signal significance $S/\sqrt{B}=2$
for $\int{\mathscr L}dt=137$ fb$^{-1}$ (red) and $\int{\mathscr L}dt=3000$ fb$^{-1}$ (blue) at $\sqrt{s}=13$ TeV, respectively.
We find that the upper bound on the FCNC coupling $g_{ut}$ could reach
about $0.6 (1.3)$ for $m_{Z'}=2 (5)$ TeV and $\int{\mathscr L}dt=137$ fb$^{-1}$, respectively.
For $\int{\mathscr L}dt=3000$ fb$^{-1}$, the upper bound could be reduced by 
a factor of about $0.6$.

In the triple top quark production, it is difficult to reconstruct all top quarks from the decay products. Therefore, in experiments, 
signal events are chosen by taking into account the decay channels,
where $tt$ ($\bar{t}\bar{t}$) decay semileptonically
and the remaining $\bar{t}$ ($t$) decays hadronically~\cite{ATLAStri}.
Then, the signals contain a pair of same-sign leptons and several $b$ or light quark jets.
We simulate the signal events by taking into account the top quark decays,
$tt\bar{t}\rightarrow bl^+\nu_l bl^+\nu_l \bar{b} j j$ 
and $t\bar{t} \bar{t} \rightarrow \bar{b}l^- \bar{\nu_l} \bar{b}l^- \bar{\nu_l} b j j$ ($l=e, \mu$), at the 13 TeV LHC.
We follow the ATLAS Collaboration for the optimal signal region (Rpc2L1bH)~\cite{ATLAStri} and require following conditions:
	
\begin{enumerate}[(i)]
\item at least one same-sign lepton pair with each lepton $p_T$ larger than $20$ GeV,
\item at least six jets with the jet $p_T$ larger than $20$ GeV,
\item at least one $b$-tagged jet with the $b$-jet $p_T$ larger than $20$ GeV,
\item $\slashed{E}_{T}$ lager than $250$ GeV,
\item $\slashed{E}_{T}/m_{\rm{eff}}$ larger than $0.2$.
\end{enumerate} 
	
We find that about $0.3\%$ of the signal events survives after requiring cuts listed above and the ATLAS Collaboration shows that 
$9.8$ events of the SM background remained after the selection for  $\int{\mathscr L}dt=36.1$ fb$^{-1}$.
For other integrated luminosities $\int{\mathscr L}dt$,
we simply 
rescale the expected number of SM background events $n_b=9.8 \times  (\int{\mathscr L}dt / 36.1)$.
	
\begin{figure}[tb]
\begin{center}
\includegraphics[width=75mm]{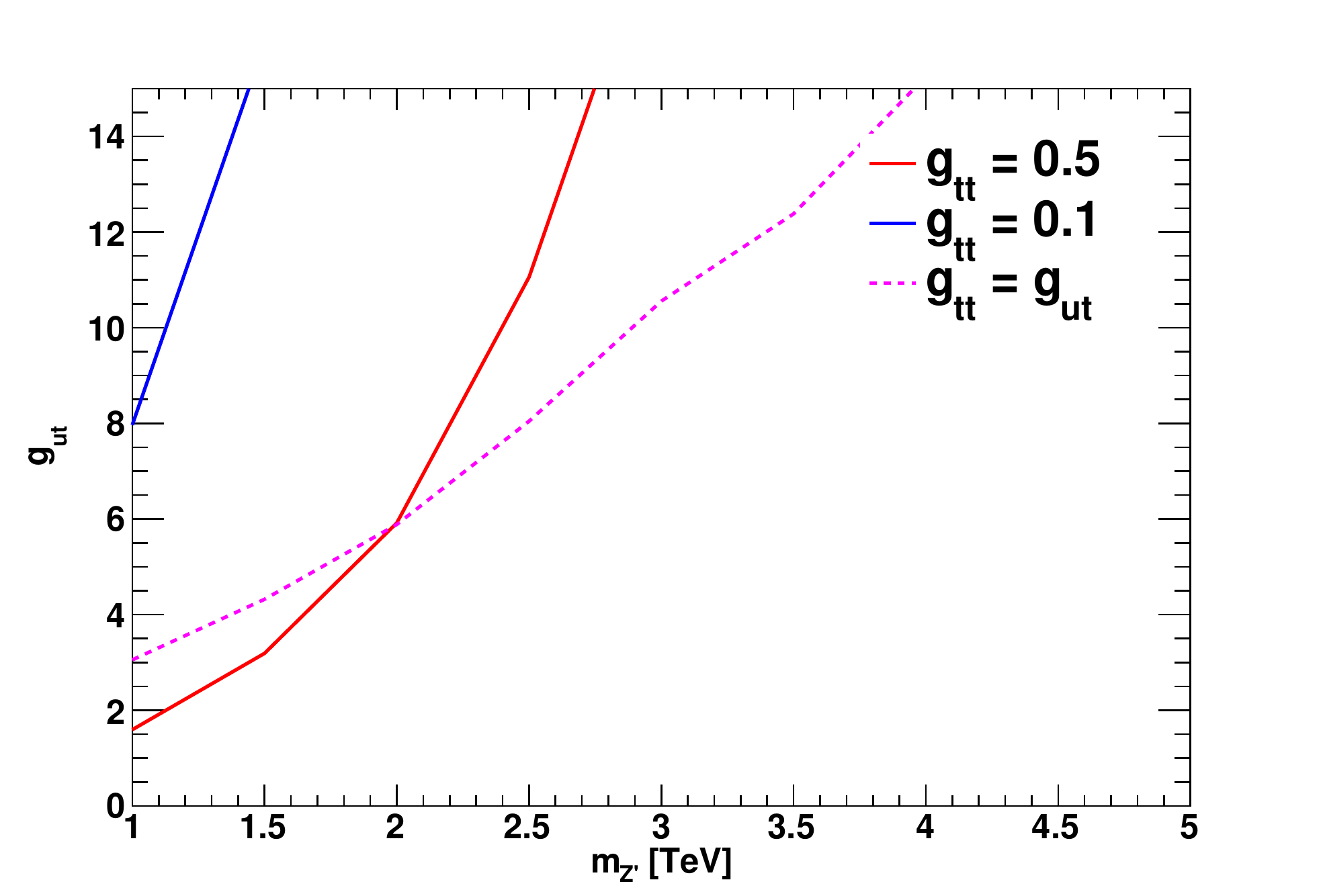} 
\includegraphics[width=75mm]{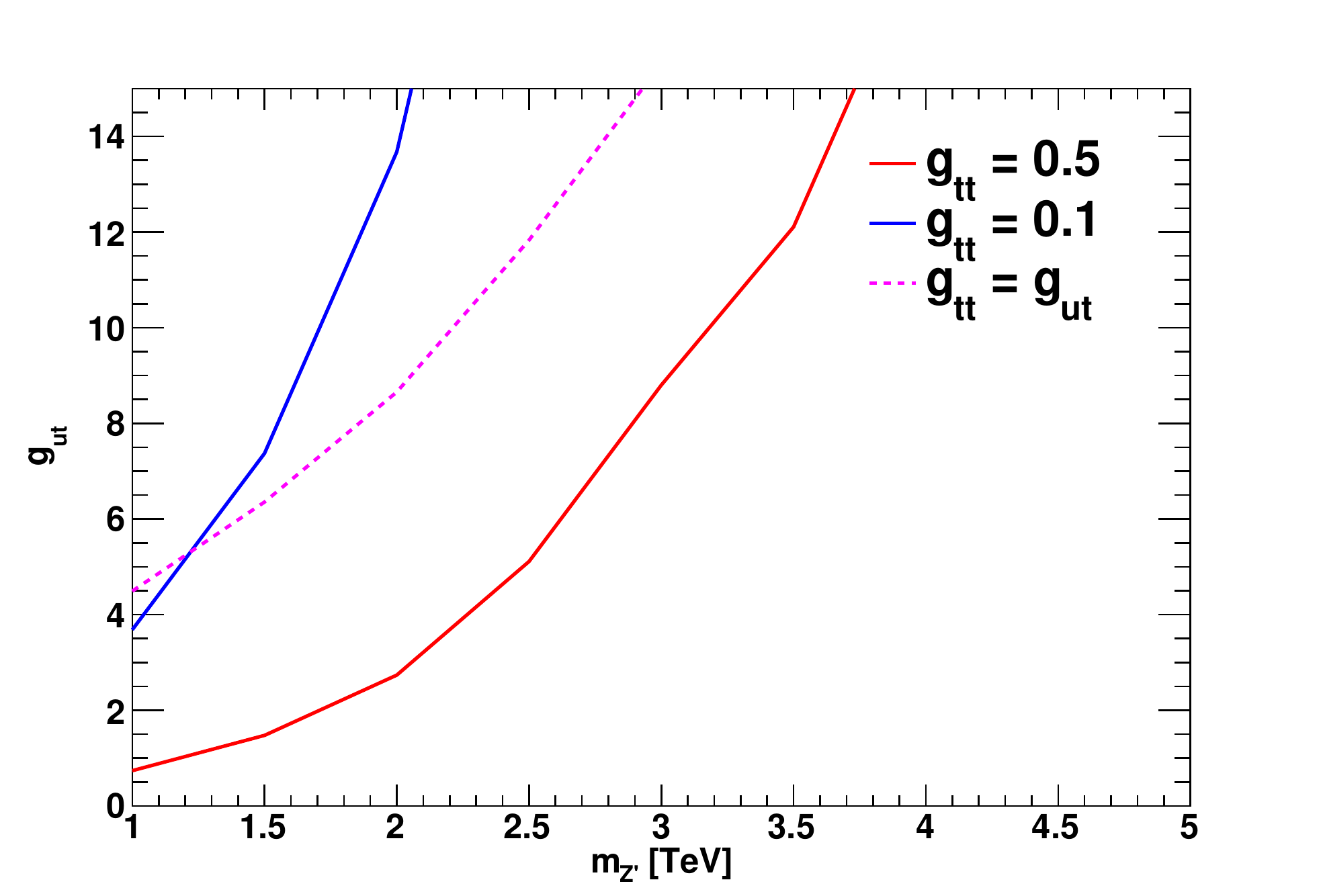}
\caption{ The contour plots of $g_{ut}$ vs $m_{Z'}$ (in TeV) for
signal significance $S/\sqrt{B}=2$ at the production level
for $\int{\mathscr L}dt=137$ fb$^{-1}$ (left) and $\int{\mathscr L}dt=3000$ fb$^{-1}$ (right) at $\sqrt{s}=13$ TeV.
The red, blue, and purple lines correspond to the cases 
of $g_{tt}=0.5$, $g_{tt}=0.1$, and $g_{ut}=g_{tt}$, respectively.
}
\label{fig:tripleATLAS}
\end{center}
\end{figure}

In Fig.~\ref{fig:tripleATLAS}, we depict the contour plots of $g_{ut}$ vs $m_{Z'}$ (in TeV) for
signal significance $S/\sqrt{B}=2$ at production level
for $\int{\mathscr L}dt=137$ fb$^{-1}$ (left) and $\int{\mathscr L}dt=3000$ fb$^{-1}$ (right) at $\sqrt{s}=13$ TeV, respectively. 
The red, blue, and purple lines correspond to the cases 
of $g_{tt}=0.5$, $g_{tt}=0.1$, and $g_{ut}=g_{tt}$, respectively.
As shown in Fig.~\ref{fig:tripleATLAS}, the upper bounds on $g_{ut}$
are greater than $O(1)$ over the whole range of $m_{Z'} \ge 1$\, TeV. 	
For $m_{Z'}=2$ TeV and $g_{tt}=0.5$, the bound could reach $5.92$ ($2.74$) for the integrated luminosity
 $\int{\mathscr L}dt=137$ ($3000$) fb$^{-1}$.
The obtained bounds on couplings at the detector level are much larger than those at the parton level in the triple top quark production.
The difference stems from the estimation of SM backgrounds.
Since the top quarks cannot be fully reconstructed in the multitop production, the signal events are selected by 
a pair of same-sign leptons with at least one $b$-tagged jet and 
a few jets in the triple top quark production~\cite{ATLAStri}.
Then, the main SM backgrounds are the $t\bar{t}+W(Z,\gamma,H)$ channels, which
have much larger cross sections than the rare multitop quark production.

Comparing the result for the upper bound on $g_{ut}$ at the production level, we find that the same-sign top quark pair production 
could provide a much more stringent bound than the triple top quark production over the whole range of $m_{Z'}\ge 1$\, TeV. 
This conclusion is apparently distinguished from the expectation
at the parton-level analysis.
The discrepancy is originated mainly from the estimation of the SM backgrounds.
In the same-sign top quark pair production, the SM backgrounds are
not so different for both parton- and detector-level analyses.
However, in the triple top quark production, the SM backgrounds 
at the detector level are much larger.

\section{Summary}
\label{sec;summary}%

In this work, we have considered a $Z'$ model in which
the $Z'$ boson couples only to $u$ and $t$ quarks.
The model could be constructed with an extra $U(1)'$ symmetry, under which only 
right-handed up-type quarks are charged.
The $Z'$ boson can be leptophobic if the mixing between
the $Z'$ boson and SM gauge bosons is negligible.
After symmetry breaking, the right-handed up-type quarks mix with each other, and the top FCNCs mediated by the $Z'$ boson can be generated at tree level.

The $Z'$ boson can be searched for through FCNC processes. In this work,
we have considered several top FCNC processes at the LHC: the same-sign top quark
pair production, single top quark production, radiative $Z'$ production,
and triple top quark production.
The first process depends on the coupling $g_{ut}$, while
$g_{uu}$ and/or $g_{tt}$ as well as $g_{ut}$ are involved in the other processes.
Therefore, such processes cannot be available for the $Z'$ search
if $g_{uu}$ and/or $g_{tt}$ are negligible.
We find that among the top FCNC processes the same-sign top quark pair production could be the most capable of yielding the best bound on the $Z'$ boson and top FCNCs induced
by the $Z'$ boson for $g_{tt}\le 0.68$ 
if only statistical uncertainties at the parton level
	are taken into account.
However, for $g_{tt}\ge 0.68$, the triple top quark production 
might provide stronger bounds on the top FCNC coupling
than the same-sign top quark pair production.

On the other hand, the $Z'$ boson can be searched for through
non-FCNC processes:
the dijet production, top quark pair production, and four top quark 
production. Only $g_{uu}$ is involved in the dijet production,
whereas $g_{ut}$ and $g_{tt}$ are also involved in the other two processes.
In the dijet production, we find that the region $g_{uu}\lesssim 0.5$
is allowed for $m_{Z'}\ge 1$ TeV and the bound could be enlarged to $0.7$
for $m_{Z'}\gtrsim 3.3$ TeV. In the top quark pair production and four top
quark production,
small $Z'$ mass and large $g_{tt}$ coupling are required to probe
the $Z'$ boson. We note that in the case of $g_{uu}\ll 1$
it may be impossible to probe the $Z'$ boson in the dijet or
top quark pair production. 
In this case, the same-sign top quark pair production or triple
top quark production might be the best probe to the $Z'$ boson
for $m_{Z'}\ge 1$ TeV.

Finally, we performed numerical analyses at detector level
for the same-sign top quark pair production and triple top quark
production by taking into account the results at the parton level.
We find that, unlike the analysis at the parton level, 
the same-sign top quark pair production provides much more stringent
bound on the FCNC coupling $g_{ut}$. The bound could reach
about $0.6 (1.3)$ for $m_{Z'}=2 (5)$ TeV and $\int{\mathscr L}dt=137$\, fb$^{-1}$.
To get complementary bound on $g_{ut}$ in the triple top quark production,
it would be necessary to reconstruct multitop quarks and 
remove the main SM backgrounds in current experiments.

\section*{Acknowledgments}
This work is supported in part by the National Research Foundation
of Korea (NRF) grant funded by the Korea government (MSIT),
Grants No. NRF-2017R1A2B4011946 (C.Y.), No. NRF-2017R1E1A1A01074699 (J.L.)
and No. NRF-2020R1A2C3009918 (S.C.).
The work of P.K. is supported in part by KIAS Individual Grant (Grant No. PG021403) 
at Korea Institute for Advanced Study and by 
National Research Foundation of Korea (NRF) Grant No. NRF-2019R1A2C3005009, 
funded by the Korea government (MSIT).
The work of C.Y. is also supported in part by a Korea University Grant.
The work of Y.O. is supported in part by the Grant-in-Aid for Scientific Research 
  from the Ministry of Education, Culture, Sports, 
 Science and Technology in Japan, Grants No. 19H04614, No. 19H05101 and No. 19K03867. 




\end{document}